\documentclass{aa}  

\usepackage{graphicx}
\usepackage{txfonts}
\usepackage{lipsum}
\usepackage{subcaption}         
\usepackage{lscape}             
\usepackage{placeins}           
                                
\usepackage{xcolor}

\begin{document}

\title{The dust attenuation scaling relation in the TNG50 simulation}

%

   \author{P.~Cataldi \inst{1}\fnmsep\thanks{Corresponding author:pcataldi@astro.unam.mx}
        \and X.~Z.~Zheng \inst{3}
        \and A.~Katsianis \inst{4}
        \and S.~Schulz \inst{5}
        \and M.~Qiao \inst{6}
        \and J.~Qin \inst{7}
        \and Z.~Lyu \inst{8,9} 
        }

   \institute{Universidad Nacional Autónoma de México, Instituto de Astronomía, A.P. 70-264, 04510, Ciudad de México, México
   \and Tsung-Dao Lee Institute and State Key Laboratory of Dark Matter Physics, Shanghai Jiao Tong University, Shanghai 201210, China
   \and School of Physics and Astronomy, Sun Yat-sen University, Zhuhai Campus, 2 Daxue Road, Xiangzhou District, Zhuhai, China
   \and Independent Researcher, Zurich, Switzerland
   \and School of Physics and Electronic Engineering, Jining University, Jining 273155, China 
   \and Hainan Micro-city Future School, Chengmai County, Hainan, 571900, China
   \and Purple Mountain Observatory, Chinese Academy of Sciences, 10 Yuanhua Road, Nanjing 210023, China
   \and School of Astronomy and Space Sciences, University of Science and Technology of China, Hefei 230026, China
   }
    \date{Received 1 May 2025 / Accepted 23 November 2026}
    
    \abstract
    {The infrared excess (IRX $\equiv L_{\mathrm{IR}}/L_{\mathrm{UV}}$) of star-forming galaxies (SFGs) is observed to follow a tight scaling relation with star formation rate, gas-phase metallicity, and galaxy size up to $z \sim 3$, providing an important benchmark for galaxy formation models.}
    {We test whether SFGs in the TNG50 simulation, post-processed with the radiative transfer code \textsc{skirt}, reproduce the universal IRX relation, which links IRX to gas-phase metallicity, infrared luminosity, size and inclination, and identify the origin of any offsets.}
    {We analyse a sample of 7280 SFGs selected from the TNG50 simulation over $0 \leq z \leq 4$, and compare the predicted IRX--galaxy parameter relations with observational constraints, and with the EAGLE simulation. We test different dust-to-metal (DTM) ratios and dust types, correct individual galaxy properties to the observed scaling relations, and decompose the offsets into the contributions of each term of the relation.}
    {We find that the TNG50-SFGs are close to the universal IRX relation at $0.5 \leq z \leq 1$, but lie above it at $z=0$ ($\approx+0.5$~dex) and below it at $z\geq2$ ($\approx-0.25$~dex). The simulated galaxies follow their own relation of the same form, with a shallower dependence on $L_\mathrm{IR}$ that is independent of metallicity. At $z\approx0$, TNG50-SFGs are $\approx1$~dex fainter in $L_\mathrm{IR}$ than the local galaxies used to calibrate the universal IRX relation, which must therefore be extrapolated; at the same stellar mass and $L_\mathrm{IR}$, simulated and observed IRX agree within 0.04~dex. At $z\geq2$, TNG50-SFGs are less obscured than the relation predicts, an offset that largely disappears if their metallicities are lowered by $\approx0.2$~dex.}
    {Varying the DTM ratio and dust composition shifts the normalisation of the offset but does not resolve its redshift dependence: a redshift-dependent DTM ratio could reduce the offset at high redshifts, but not at low redshifts. Because the low-redshift discrepancy arises largely outside the $L_\mathrm{IR}$ and metallicity range over which the universal IRX relation was calibrated, resolving it will require both more physically motivated dust models and observational constraints on IR-faint galaxies.}
    

\maketitle

\section{Introduction}
\label{sect:Introduction}

In galaxy formation and evolution, many galaxy properties are interrelated and mutually influence one another, giving rise to tight empirical relations. For example, the infrared excess (IRX), star formation rate (SFR), gas-phase metallicity, and galaxy size all follow well-defined relationships \citep[e.g.,][]{Whitaker2014,Popesso2023,Sanders2023,Suess2019a}. Although dust accounts for only $\sim1\%$ of the interstellar medium (ISM) mass in star-forming galaxies (SFGs), it plays a large role in shaping their spectral energy distributions (SEDs) as it absorbs and scatters UV–optical photons, then re-emits the absorbed energy in the infrared \citep[i.e., IR,][]{Galliano2018, Hahn2022}.

This process significantly biases the inferred galaxy properties if not properly accounted for \citep{Calzetti2001, Conroy2013, Trayford2020}. At $z>3$, SFRs are commonly inferred from UV emission, which traces the presence of young, massive stars and therefore recent star formation activity \citep{Kennicutt1998}. However, dust attenuation effects have to be considered to estimate the intrinsic SFRs \citep[e.g.,][]{Katsianis2021}. The dust content of a galaxy depends on its gas reservoir and gas-to-dust ratio, the latter being closely linked to the gas-phase metallicity \citep{Trayford2017, Galliano2018, DeVis2019, Trcka2020, Casasola2020}. Observationally, attenuated stellar emission and dust thermal IR radiation can be measured directly; however, the total dust mass is inferred indirectly and remains model-dependent. 

Dust attenuation quantifies the fraction of stellar radiation that is absorbed and scattered out of (and into) the line of sight by dust \citep{Salim2020}. In practice, it is commonly characterised by the infrared excess \citep{Heckman1998, Meurer1999, Heinis2013, Bourne2017}. The strength of dust attenuation is primarily regulated by the star formation surface density and the effective dust column density, and is therefore tightly connected to star formation, metallicity, and galaxy compactness \citep{Garn2010, Xiao2012, Battisti2016, Katsianis2017a, Zahid2017, Shapley2022, Qin2024}.

Interestingly, \citet{Qin2019} showed that IRX can be expressed as a joint power-law function of gas-phase metallicity, IR luminosity ($L_{\mathrm{ IR}}$), half-light radius ($R_{\mathrm{e}}$), and galaxy inclination ($b/a$), of the form ${\mathrm{IRX}} = 10^\alpha\, L_{\mathrm{IR}}^\beta\, R_{\mathrm{e}}^{-\gamma}\,(b/a)^{-\delta}$, where the power-law indices depend explicitly on metallicity. This formulation establishes a unified IRX relation for SFGs up to $z \sim 3$, encapsulating how dust attenuation is governed by the interplay between metal enrichment, star formation activity, galaxy compactness, and viewing angle. In this framework, gas-phase metallicity emerges as the key parameter modulating the relative importance of the other quantities. As metallicity decreases, the power-law slopes systematically flatten, leading to an IRX behaviour that departs markedly from expectations based on the Kennicutt--Schmidt law \citep{Kennicutt1998} combined with a constant dust-to-metal (DTM) ratio \citep{Wuyts2011, Genzel2013, Qin2019}.

The above result provides a natural explanation for the apparent lack of evolution in the $M_{\ast}$--IRX relation for SFGs with $M_{\ast} < 10^{10.5}\,{\mathrm {M_\odot}}$ over $0<z<3$. The universal IRX relation highlights that dust attenuation is jointly regulated by star formation, ISM metallicity, and galaxy structure, making it a fundamental benchmark for theoretical models. 

However, reproducing this relation has proven challenging for cosmological simulations at higher redshift. For example, while EAGLE successfully matches several dust scaling relations, it struggles to recover the full universal IRX relation \citep{ManQiao2024}. This tension has been attributed to simplified dust prescriptions at high redshift and/or to limitations in how simulations model star formation and chemical enrichment in the early Universe. Recent observations also suggest that deviations from the universal relation can arise from differences in star--dust geometry, with galaxy compactness and inclination playing an important role in shaping the IRX scatter \citep{Lyu2026}.

We use the TNG50 cosmological hydrodynamical simulation, combined with the radiative transfer code \textsc{skirt}, to investigate the theoretically predicted IRX in relation to other galaxy properties, and to assess whether the observed universal IRX relation can be reproduced within this framework. Following a similar methodology as \citet{ManQiao2024}, we validate TNG50 predictions using this time the catalogue of \citet{Schulz2020}.

This paper is organised as follows. In Sect.~\ref{sect:methods}, we describe the TNG50 simulation, sample selection, and galaxy parameter estimation. In Sect.~\ref{sect:results}, we present the predicted IRX relation, compare it with EAGLE-SFGs and observations, test the dust post-processing, and decompose the offsets into the contributions of the individual terms of the relation. Sect.~\ref{sect:discussion} contains discussion and possible interpretation of our findings. Finally, we summarise our main results in Sect.~\ref{sect:conclusions}.

\section{Methodology}
\label{sect:methods}

\subsection{Sample selection}

Our analysis is based on the SFGs from the TNG50 simulation (hereafter, TNG50-SFGs) from the \textsc{IllustrisTNG} project \citep{Nelson2019,Pillepich2018}. The classification of SFGs is done as in \citet{Pillepich2019}. The star-forming main sequence (SFMS) is defined as the median specific star formation rate (sSFR) in 0.2-dex stellar-mass bins, and galaxies are classified according to their offset from this relation: SFGs have $\Delta\log_{10}\mathrm{sSFR} > -0.5$, while green-valley and quenched systems lie at $-1.0 < \Delta\log_{10}\mathrm{sSFR} < -0.5$ and $\leq -1.0$, respectively.

We make use of the synthetic galaxy catalogue presented by \citet{Schulz2020}, who post-processed TNG50 galaxies with the radiative transfer code \textsc{skirt} to derive UV and IR luminosities. In this post-processing, each galaxy is reconstructed on its native Voronoi mesh \citep{Springel2010, Camps2013}, extracting a cubic region of side length $15\, R^{\star}_{\mathrm{hm}}$ to capture the diffuse dust component, where $R^{\star}_{\mathrm{hm}}$ represents the half-mass radius of the distribution of stellar particles. Since TNG50 does not explicitly model dust, the dust density is assumed to trace the galaxies’ gas-phase metals. Dust is assigned to gas cells that are either star-forming (SFR > 0) or have temperatures below $T_{\mathrm{gas}} < 7.5\times10^4 \,\mathrm{K}$, such that

\begin{equation}
\rho_{\mathrm{dust}} =
\begin{cases}
f\,\rho_{\mathrm{gas}}\,Z & \text{if } (T_{\mathrm{gas}} < 7.5\times10^4\,{\mathrm{K}}) \ \text{or}\ ({\mathrm{SFR}}_{\mathrm{gas}} > 0) \\
0 & \text{if } (T_{\mathrm{gas}} \geq 7.5\times10^4\,{\mathrm{K}}) \ \text{and}\ ({\mathrm{SFR}}_{\mathrm{gas}} = 0)
\end{cases}
\label{eq:eq1}
\end{equation}

\noindent where $\rho_{\mathrm{gas}}$ and $Z$ are the gas density and metallicity, respectively. This prescription excludes hot, non–star-forming gas, where dust is expected to be efficiently destroyed by thermal sputtering \citep{Guhathakurta1989}, while allowing dust to persist in cooler or star-forming phases, including outflows and the extended circumgalactic medium \citep[e.g., dusty outflows,][]{Li2025}. We adopt a constant DTM ratio of $f_{\mathrm{dust}} = 0.3$, regardless of galaxy type or redshift, consistent with previous studies \citep[e.g.,][]{Camps2016, Camps2018, Rodriguez-Gomez2019}. 

To accurately model the attenuation of TNG50 galaxy spectra by dust in the ISM, radiative transfer calculations were performed by \citet{Schulz2020} using the three-dimensional dust radiative transfer code \textsc{skirt} \citep{Baes2011,Camps2015,Baes2020}, in which the Milky Way (MW)-like dust mixture model was adopted \citep{Zubko2004}. Stellar populations younger than 10\, Myr include birth-cloud attenuation via \textsc{mappings-iii} \citep{Groves2008}, while older stars are attenuated only by the diffuse ISM. We refer the readers to \citet{Schulz2020} for more details. This prescription is adopted as our fiducial model; we additionally test the sensitivity of our results to the DTM ratio and dust composition (MW, LMC, and SMC) for a representative subsample of galaxies, as described in Sect.~\ref{sec:f_dust_var}.

The catalogue presented in \citet{Schulz2020} follows an almost identical procedure to that of \citet{Camps2018}, implemented for EAGLE-SFGs. The only difference lies in the adopted side length of the region used for the diffuse dust component (a variable $15 \, R^{\star}_{\mathrm{hm}}$ versus a fixed 30 pkpc, respectively), which has no significant impact on the IRX modelling that is the focus of this work. The methodology is also closely aligned with that of the public photometric database presented in \citet{Trcka2022}. Although the latter adopts a lower DTM ratio ($f_{\mathrm{dust}} = 0.2$) and benefits from a larger statistical sample ($\sim 14{,}000$ galaxies), it is limited to low-redshift systems ($z \leq 0.1$), which restricts its applicability and the evolution of the relationships we focus on in this work.

\begin{figure}
    \centering    
    \includegraphics[width=0.8\columnwidth]{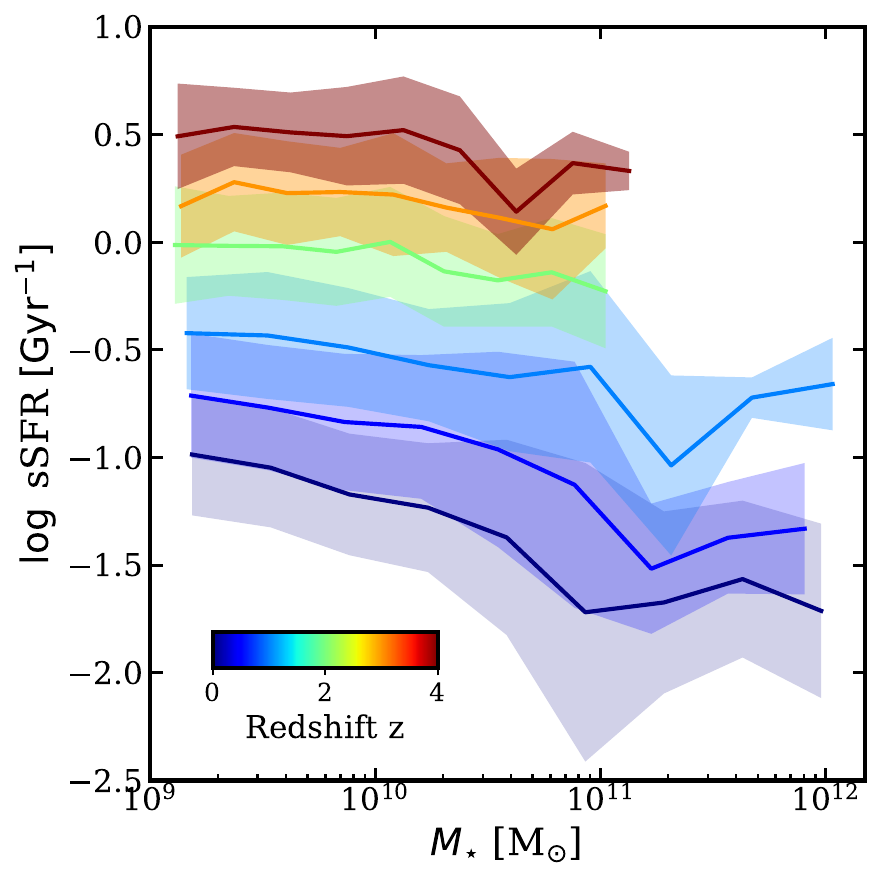}
    \caption{TNG50-SFGs median $M_{\ast}$--$\mathrm{sSFR}$ relation, colour-coded by redshift. The shaded regions indicate the $\pm1\sigma$. As we increase redshift, fewer galaxies are found in the selected stellar mass range (> $10^{9} \mathrm{M_{\odot}}$), while the average sSFR gradually increases.}
    \label{fig:fig1}
\end{figure}

Using the TNG50—a high-resolution simulation post-processed by a radiative transfer code—is ideal for a direct comparison to the results from \citet{ManQiao2024}, and the extension of this kind of analysis to one of the most popular galaxy formation models (\textsc{IllustrisTNG}). Fig.~\ref{fig:fig1} shows the resulting $M_{\star}$–sSFR relation, and Table~\ref{tab:tab1} lists the number of selected galaxies at each redshift.

\begin{table}
\caption{Summary of the TNG50 galaxy sample used in this work.}
\label{tab:tab1}
\centering
\begin{tabular}{c c c}
\hline\hline
Redshift & \multicolumn{2}{c}{\# TNG50 galaxies} \\
\cline{2-3}
$z$ & $M_\star > 10^9\,\mathrm{M_{\odot}}$ & SF and $M_\star > 10^9\,\mathrm{M_{\odot}}$ \\
\hline
0.0 & 2524 & 1790 \\
0.5 & 2366 & 1767 \\
1.0 & 2090 & 1710 \\
2.0 & 1272 & 1149 \\
3.0 & 652  & 620  \\
4.0 & 249  & 244  \\
\hline
\end{tabular}
\tablefoot{Columns list the redshift, the total number of galaxies with $M_\star > 10^9\,\mathrm{M_{\odot}}$, and the number of selected SFGs according to \citet{Pillepich2019}. Although the total number of galaxies decreases towards higher redshift, the fraction classified as SFGs increases.}
\end{table}

\subsection{Galaxy parameters}
\label{sec:parameters}
The galaxy parameters—including stellar mass, SFR, absolute abundances of hydrogen and oxygen of star-forming gas, and broadband fluxes—are all derived from a spherical aperture designed to encompass most of the galaxy light. Different observational studies adopt a variety of aperture definitions \citep{Graham2005,Hill2011}, and identifying their numerical analogues is not straightforward \citep[see discussion in][]{Trcka2022}. In this work, we adopt an aperture of $2 R^{\star}_{\mathrm{hm}}$, in line with previous studies \citep{Rodriguez-Gomez2019, Popping2022}. \footnote{In \citet{ManQiao2024}, galaxy properties are measured within a fixed spherical aperture of 30\,pkpc. This difference has no meaningful impact on our results, as both $2 R^{\star}_{\mathrm{hm}}$ and 30\,pkpc typically encompass the full stellar extent of the galaxy.}

We focus on the set of parameters used to construct the universal IRX relation following \citet{Qin2019}. In particular, we convert the half-mass radii provided by the simulation ($R^{\star}_{\mathrm{hm}}$) into half-light radii ($R_{\mathrm{e}}$). For galaxies at $0 \leq z \leq 2$, we adopt the procedure described in \citet{ManQiao2024}, which is based on the observational constraints of \citet{Suess2019a, Suess2019b} and has been validated against mock-image analyses such as \citet{deGraaff2022}. We refer the reader to \citet{ManQiao2024} for a detailed description of this methodology.

We extend this approach to higher redshifts, up to $z \leq 4$. However, in the regime $3 \leq z \leq 4$, the conversion becomes significantly more uncertain due to the strong dependence of mass-to-light ratios on the observed band and wavelength, which probe different stellar populations with distinct spatial distributions \citep{Varadaraj2024, Clausen2025, Xin2026}. For this redshift range, we therefore adopt the mass-to-light ratio conversion proposed by \citet{Costantin2023} for the band F200W, derived from mock catalogues of SFGs in TNG50.

The output of the \textsc{skirt} runs consists of SEDs of each galaxy, provided as flux density $F_\lambda = \lambda f_\lambda$ (in $\mathrm{W/m^2}$) as a function of wavelength $\lambda$ (in $\mathrm{\mu m}$). The infrared excess is defined as 

\begin{equation}
\mathrm{IRX} = \frac{L_{\mathrm{IR}}}{L_{\mathrm{UV}}},
\label{eq:eq2}
\end{equation}

\noindent where $L_{\mathrm{IR}}$ is obtained by integrating the spectral luminosity over $8$–$1000\,\mu$m, and $L_{\mathrm{UV}}$ is the luminosity at 1600\,\AA. For fixed dust properties and geometry, greater dust attenuation yields higher IRX values. We note that \citet{Schulz2020} find that their synthetic catalogue reproduces well the observationally derived IRX–$\beta^{\mathrm{UV}}$ relations at $z \lesssim 1$ (where $\beta^{\mathrm{UV}}$ is the rest-frame UV continuum slope, assuming a power-law $f_{\lambda} \propto \lambda^ {\beta^{\mathrm{UV}}}$). 

Figure~\ref{fig:figA1} shows the relationship between IRX and the main physical quantities for our sample, compared with the EAGLE-SFGs analysed in \citet{ManQiao2024} (shown as red contours) and the local SFG sample from \citet{Qin2019} at $z\sim0$ (shown as black contours), from which the universal IRX relation was originally calibrated and shown to hold for stacked samples up to $z\approx3$. We observe that $M_{\star}$, SFR, Z, and $L_{\mathrm{IR}}$ correlate positively with IRX. A complete summary of the medians, minimum, and maximum values of these parameters across all redshift bins and samples is provided in Table~\ref{tab:sample_properties}. We note that the samples present different regimes: the EAGLE-SFGs extend up to $z<2.5$ and include lower-mass galaxies ($M_{\star} > 10^{8.5} \, \mathrm{M_{\odot}}$), whereas the TNG50-SFGs reach higher redshifts ($z<4$) but adopt a higher mass cut ($M_{\star} > 10^{9} \,\mathrm{M_{\odot}}$). Furthermore, the observational sample of \citet{Qin2019} (drawn from SDSS, GALEX, and WISE) has higher SFRs—consistent with the SFR offsets discussed for TNG50 in \citet{Donnari2019}—, lower gas-phase metallicities, and systematically higher IR luminosities. Despite these differences, the overall trends in both simulations are broadly consistent. However, when comparing the two simulations at fixed metallicity and infrared luminosity $L_{\mathrm{IR}}$, TNG50-SFGs exhibit systematically higher IRX values than EAGLE-SFGs, consistent with a higher level of dust attenuation in TNG50. Conversely, the observational sample displays the highest IRX values overall, as expected from its higher $L_\mathrm{IR}$; at matched $M_\ast$ and $L_\mathrm{IR}$, the IRX of TNG50 agrees with the observed one (Sect.~\ref{sec:z0}).

Finally, the half-light radius, $R_{\mathrm{e}}$, distribution in TNG50-SFGs is noticeably broader, extending down to $\sim 0.2 \,\mathrm{kpc}$, whereas EAGLE-SFGs reach minimum sizes of $\sim 0.6 \, \mathrm{kpc}$.

\section{Results}
\label{sect:results}

The universal IRX relation introduced by \citet{Qin2019} expresses dust attenuation as a coupled power-law dependence on gas-phase metallicity, infrared luminosity, half-light radius, and axial ratio:

\begin{equation}
\mathrm{IRX} = 10^\alpha\, \left ( \dfrac{L_{\mathrm{IR}}}{10^{10}\mathrm{L_{\odot}}}\right )^\beta\, \left ( \dfrac{R_{\mathrm{e}} }{1 \,\mathrm{kpc}}\right )^{-\gamma}\, (b/a)^{-\delta},
\label{eq:eq3}
\end{equation}

\noindent where

\begin{equation}
\begin{aligned}
\alpha &= 1.07\log (Z/Z_\odot) + 0.95, \\
\beta &= 0.91\log (Z/Z_\odot) + 0.64, \\
\gamma &= 1.15\log (Z/Z_\odot) + 0.80, \\
\delta &= 1.43\log (Z/Z_\odot) + 0.94.
\end{aligned}
\label{eq:eqIndex}
\end{equation}

\begin{figure*}
    \centering 
    \includegraphics[width=\textwidth]{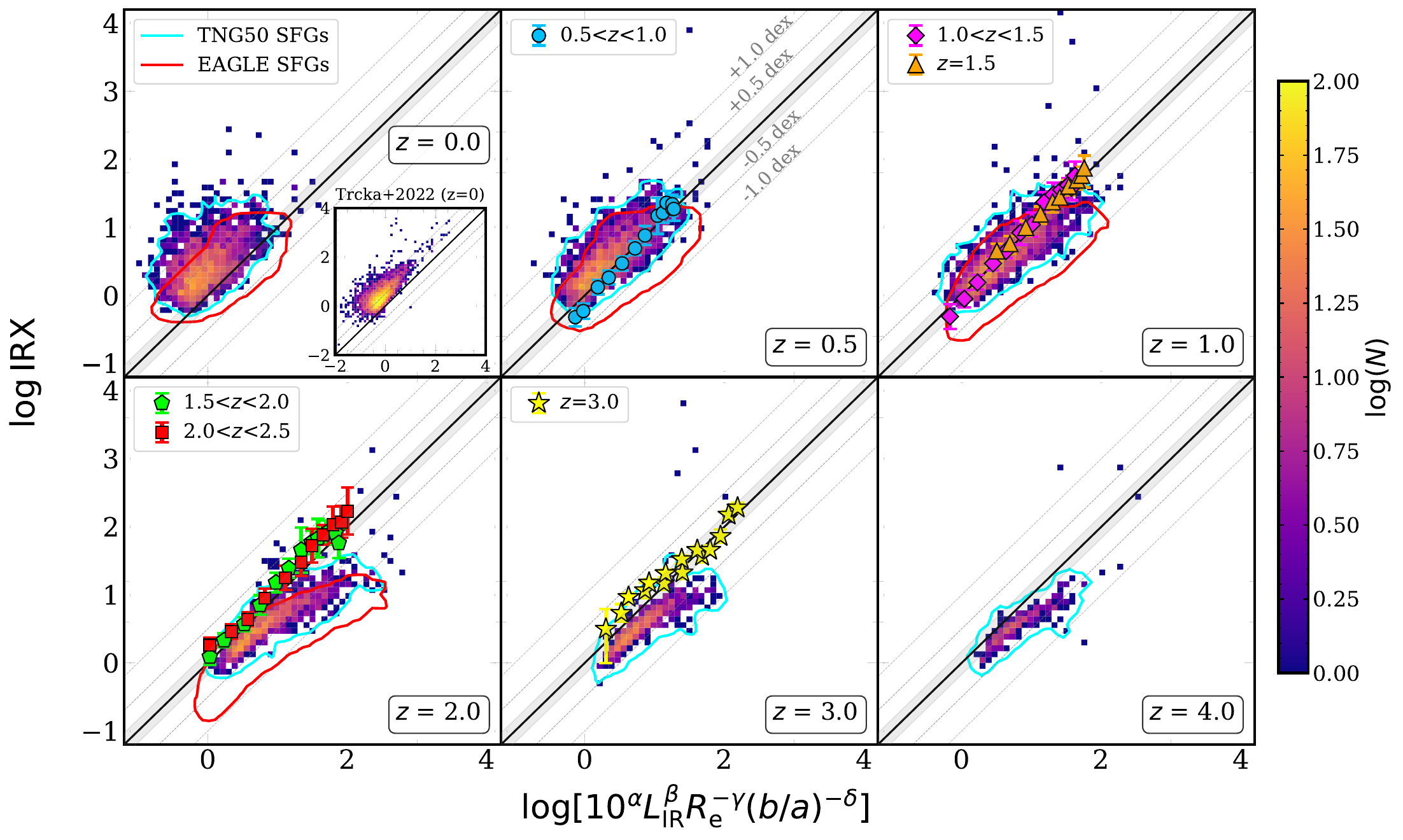}
    \caption{Distributions of the recovered IRX are shown in comparison with the universal IRX relation obtained using Eq.~\ref{eq:eq3} across six redshifts for the TNG50-SFGs sample. The grey shaded region represents the $\pm\sigma$ dispersion reported by \citet{Qin2019}. Data points with error bars correspond to average SFGs in different stellar mass bins at each redshift, taken from \protect\citet{Whitaker2014} for $0.5<z<2.5$, \protect\citet{Heinis2014} for $z=1.5$ and $z=3.0$, and \protect\citet{AlvarezMarquez2016} for $z=3$. The cyan contour marks the outer outline of the distribution of SFGs, while the red contour lines indicate the distribution of the EAGLE-SFGs sample presented by \citet{ManQiao2024}. In the $z=0$ subpanels, the inset shows the IRX distribution of TNG50-SFGs from \citet{Trcka2022}. We find that TNG50-SFGs broadly follow the observed relation at intermediate redshifts ($z=0.5$ and $z=1.0$), whereas noticeable deviations are observed at $z=0$ and at $z \geq 2$.}
    \label{fig:fig2}
\end{figure*}

\noindent Here, $\log (Z/Z_{\odot}) = 12 + \log(\mathrm{O/H}) - 8.69$ represents the oxygen abundance relative to solar, where $Z_{\odot}$ denotes the Solar metallicity. In this formulation, IRX increases with metallicity and infrared luminosity, and decreases with galaxy size and axial ratio. Fig.~\ref{fig:fig2} compares the universal IRX relation derived from Eq.~\ref{eq:eq3} with the IRX obtained from Eq.~\ref{eq:eq2} for the sample. In each panel, the coloured map shows the TNG50-SFGs distribution for a different redshift. The cyan lines indicate the outer outline of the distribution of TNG50-SFGs. Since the IRX of each galaxy is computed for a random line of sight, we adopt $b/a=0.6$, which corresponds to the average attenuation over $4\pi$ steradians \citep[see][]{Qin2019} and therefore represents the inclination averaged over the population.

For comparison with other simulations and post-processing methodologies, we include in Fig.~\ref{fig:fig2} the results of \citet{ManQiao2024} for EAGLE-SFGs, shown as red contour lines. In addition, the inset panel at $z=0$ shows the IRX results from \citet{Trcka2022}, who use TNG50 simulated galaxies with a similar methodology to \citet{Schulz2020}, but with improved statistics owing to a lower stellar mass cut ($> 10^8 \, \mathrm{M_{\odot}}$) and a lower DTM ratio, $f_{\mathrm{dust}} = 0.2$.

To quantify the discrepancy between the recovered IRX and the universal IRX relation described by Eq.~\ref{eq:eq3}, we define the IRX residual as $\Delta \log \mathrm{IRX} = \log \mathrm{IRX}_{\mathrm{true}} - \log \mathrm{IRX}_{\mathrm{model}}$. At $z=0.5$ and $z=1.0$, the TNG50-SFGs broadly follow the universal IRX relation, albeit with a scatter of $\sim 0.5$ dex, larger than the $\sigma = 0.17$ dex reported by \citet{Qin2019}.

However, we identify a systematic offset that evolves with redshift: the universal relation underpredicts the IRX of TNG50-SFGs at $z=0$ ($\Delta \log \mathrm{IRX} \gtrapprox 0.5$ dex), agrees with it approximately at intermediate redshifts ($\Delta \log \mathrm{IRX}\sim 0$ dex), and overpredicts it at higher redshifts ($\Delta \log \mathrm{IRX} \lessapprox - 0.25$ dex).

The offset can be separated into two distinct regimes: a low-$z$ component and a high-$z$ component. At $z=0$, the deviation observed in TNG50-SFGs is approximately constant across the full IRX range, preserving the overall shape of the proposed universal IRX relation but with a systematic shift of $\sim 0.5$ dex. Notably, this offset changes little between $f_{\mathrm{dust}}=0.2$ and 0.3, the values adopted in \citet{Trcka2022} and in this work (see also Sect.~\ref{sec:f_dust_var}). Additionally, when varying the spherical aperture, as shown in Fig.~\ref{fig:figTrcka} (Appendix~\ref{sec:appendix}), the residual $\Delta \log \mathrm{IRX}$ remains largely unchanged. In both cases, somewhat surprisingly, the models at $z=0$ fail to reproduce the universal IRX relation, despite broadly matching other observational constraints, such as the IRX–$\beta^{\mathrm{UV}}$ relation \citep{Schulz2020} and the luminosity function around the knee \citep{Trcka2022}. The $L_\mathrm{UV}$--$L_\mathrm{IR}$ distributions of our sample and of \citet{Trcka2022} are very similar at $z=0$ (median $\log\mathrm{IRX}=0.39$ and 0.34, respectively), so the offset does not arise from differences between the two post-processing pipelines. Reproducing the IR luminosity function is also not necessarily in contradiction with lying above the universal relation, since the luminosity function involves the whole galaxy population, whereas the universal IRX relation was calibrated on IR-bright galaxies (Sect.~\ref{sec:calibration}).

At higher redshifts, substantial deviations from the observed universal IRX relation emerge at $z \gtrsim 2$, with a clear dependence on IRX. The discrepancy is most pronounced at the high-IRX end, corresponding predominantly to galaxies with $\log (M_\ast/\mathrm{M}_\odot) \sim 10.5$. For instance, at $z=2$, these systems exhibit IRX values up to an order of magnitude lower than those predicted from their $Z$, $L_{\mathrm{IR}}$, $R_{\mathrm{e}}$, and $b/a$ under the universal relation. A similar trend is found in the EAGLE-SFGs analysed by \citet{ManQiao2024} (see Sect.~\ref{sec:calibration} for its origin in TNG50).

At high-$z$, \citet{Schulz2020} analysing the same galaxy sample showed that offsets in locally calibrated IRX–$\beta^{\mathrm{UV}}$ relations can reach up to $\sim 0.3$ dex in IRX at fixed $\beta^{\mathrm{UV}}$, correlating with intrinsic galaxy properties such as SFR, star formation efficiency, gas metallicity, and stellar mass (see, e.g., their Fig.~6). Stronger observational constraints on the IR emission of galaxies at $z \gtrsim 2$ would help to establish whether these offsets reflect the limitations of the dust modelling or uncertainties in the galaxy properties on which the relation is built.

\subsection{Varying the DTM ratio and dust type}
\label{sec:f_dust_var}

To investigate whether the discrepancy between the TNG50-SFGs and the observed IRX scaling relation could be related to the assumptions adopted for the dust post-processing, we test the sensitivity of our results to both the DTM ratio and the dust grain composition. The fiducial model assumes a MW-type dust composition based on \citet{Zubko2004} and a DTM ratio of $f_{\mathrm{dust}}=0.3$. We select a subsample of 100 SFGs spanning the stellar-mass range of our sample at each of six redshift bins and reprocess the galaxies using different dust prescriptions.

\begin{figure}
    \centering
    \includegraphics[width=\columnwidth]{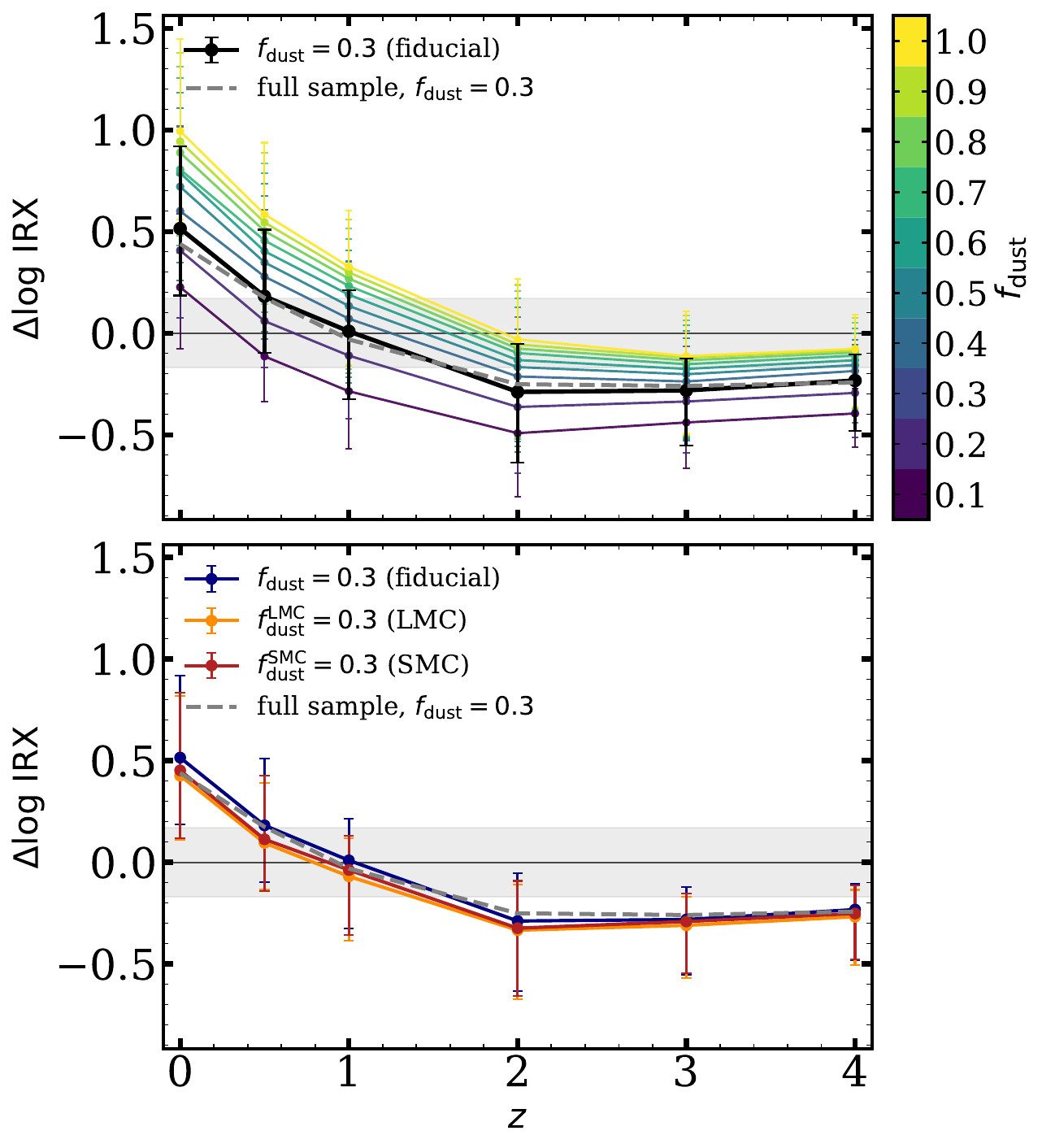}
    \caption{IRX residuals, $\Delta \log \mathrm{IRX}$, as a function of redshift, for different assumptions on the DTM ratio $f_{\mathrm{dust}}$ (top panel) and dust type (bottom panel), computed for a subsample of 100 TNG50-SFGs per snapshot. In the top panel, curves show the MW-type dust for $f_{\mathrm{dust}}=0.1$--$1.0$ in steps of $0.1$, with the fiducial value ($f_{\mathrm{dust}}=0.3$) highlighted in black. In the bottom panel, coloured curves show the MW, LMC, and SMC dust types. In both panels, points and error bars show the median and $16^{\mathrm{th}}$--$84^{\mathrm{th}}$ percentile dispersion of $\Delta \log \mathrm{IRX}$ at each snapshot, and the grey dashed line shows the median for the full TNG50-SFG sample with the fiducial MW-type dust and $f_{\mathrm{dust}}=0.3$. The horizontal black line marks $\Delta \log \mathrm{IRX}=0$ dex, and the shaded grey region indicates the global $\pm\sigma$ dispersion ($\sim0.17$ dex) reported in \citet{Qin2019}.}
    \label{fig:deltaIRX_vs_z_combined}
\end{figure}

First, we vary the DTM ratio while keeping the MW dust composition fixed. We consider $f_{\mathrm{dust}}=0.1-1.0$ in steps of 0.1. Increasing $f_{\mathrm{dust}}$ increases the amount of dust associated with the gas and therefore the dust opacity, resulting in higher $L_{\mathrm{IR}}$, and higher IRX values. Conversely, decreasing $f_{\mathrm{dust}}$ reduces the dust opacity and shifts the galaxies towards lower IRX. As shown in the top panel of Fig.~\ref{fig:deltaIRX_vs_z_combined}, varying $f_{\mathrm{dust}}$ primarily changes the normalisation of the IRX residuals, while their overall redshift dependence remains qualitatively similar.

This behaviour can also be understood in terms of Eq.~\ref{eq:eq3}. For a given TNG50 galaxy, changing $f_{\mathrm{dust}}$ changes the dust mass and therefore both $L_{\mathrm{UV}}$ and $L_{\mathrm{IR}}$. Since the absorbed UV light is re-emitted in the IR, and $L_{\mathrm{IR}}$ cannot exceed the stellar luminosity available for absorption, the measured IRX changes mainly through $L_{\mathrm{UV}}$. Between $f_{\mathrm{dust}}=0.1$ and 1, the median IRX of the subsample increases by 1.0~dex at $z=0$, of which 0.3~dex comes from the increase in $L_\mathrm{IR}$ and 0.7~dex from the decrease in $L_{\mathrm{UV}}$. At $z\geq2$, where $L_\mathrm{IR}$ is close to saturation, IRX increases by only 0.3--0.5~dex, almost entirely through $L_{\mathrm{UV}}$, while $L_\mathrm{IR}$ changes by less than 0.1~dex (not shown). The IRX predicted by Eq.~\ref{eq:eq3} responds only to the change in $L_{\mathrm{IR}}$, since $Z$, $R_{\mathrm{e}}$ and $b/a$ are set by the underlying galaxy and are not modified by the post-processing. Increasing $f_{\mathrm{dust}}$ therefore raises the measured IRX more than the predicted one, and $\Delta\log\mathrm{IRX}$ increases at all redshifts, most strongly at low redshift. This reduces the negative residuals at high redshift but increases the positive residuals at $z=0$, so no single value of $f_{\mathrm{dust}}$ removes the redshift dependence of $\Delta\log\mathrm{IRX}$.

We also investigate the effect of the assumed dust composition by replacing the fiducial MW dust with Large and Small Magellanic Cloud (LMC and SMC) dust at fixed $f_{\mathrm{dust}}=0.3$. \textsc{skirt} provides different dust models that account for variations in the grain composition and size distribution. In our fiducial model, following \citet{Camps2016} and \citet{Trayford2017}, we adopt the multi-component dust mixture of \citet{Zubko2004}, which includes graphite, silicate, and polycyclic aromatic hydrocarbon (PAH) grains with different grain-size distributions, chosen to reproduce the properties of MW-type dust. For the LMC and SMC models, we instead adopt the corresponding grain-size distributions of \citet{Weingartner2001}. These different dust compositions produce different extinction curves and therefore modify how the stellar radiation is attenuated and reprocessed into the infrared. In particular, LMC and SMC dust have a higher FUV-to-NUV extinction ratio than MW dust, with the effect being stronger for SMC dust. Consequently, changing from MW to LMC and SMC dust is expected to produce progressively stronger changes in the UV attenuation and in the IRX--$\beta^{\mathrm{UV}}$ relation, with LMC and SMC dust generally producing a shallower IRX--$\beta^{\mathrm{UV}}$ relation than MW dust.

The resulting IRX residuals are shown in the bottom panel of Fig.~\ref{fig:deltaIRX_vs_z_combined}. Despite the differences in the extinction curves, changing the dust composition produces only a relatively mild modification of $\Delta\log\mathrm{IRX}$ over the redshift range explored. Neither the LMC nor the SMC dust prescription substantially changes the overall redshift dependence of the residuals, and the systematic discrepancy between the simulated galaxies and the observed IRX relation persists for all three dust compositions. Thus, although the choice of dust grain composition can modify the detailed UV attenuation and the location of galaxies in the IRX--$\beta^{\mathrm{UV}}$ plane (see Fig.~10 of \citet{Schulz2020}), it does not by itself explain the redshift-dependent discrepancy between the TNG50-SFGs and the observed IRX scaling relation.

As shown in Fig.~\ref{fig:deltaIRX_vs_z_combined}, no single value of $f_{\mathrm{dust}}$ minimises the residuals at all redshifts: lower values reduce the offset at $z=0$ but increase it at $z\geq2$, and vice versa. We therefore retain the MW-type dust model with $f_{\mathrm{dust}}=0.3$, as adopted in previous studies, as our fiducial model for the remainder of the analysis.

Overall, these tests indicate that the discrepancy between the TNG50-SFGs and the observed IRX scaling relation cannot be explained solely by the assumed DTM ratio or dust grain composition in the post-processing. Varying $f_{\mathrm{dust}}$ shifts the residuals at all redshifts, but it does not reproduce the observed redshift dependence over the full redshift range considered. Similarly, changing the dust composition has only a mild effect on the residuals. The persistence of the redshift dependence of $\Delta\log\mathrm{IRX}$ across these dust prescriptions suggests that it originates in the properties of the simulated galaxies or in the range over which the observed relation is calibrated (Sect.~\ref{sec:calibration}).

\subsection{Correction to the simulated galaxy parameters}
\label{sec:corrections}

We investigate the scaling relations of the simulated SFGs to identify the origin of their deviations from the universal IRX relation. To this end, we compare the TNG50-SFGs with observational scaling relations for the key galaxy properties entering the \citet{Qin2019} formulation, namely $R_{\mathrm{e}}$, $Z$, and $L_{\mathrm{IR}}$.

\begin{figure}
    \centering
    \includegraphics[width=\columnwidth]{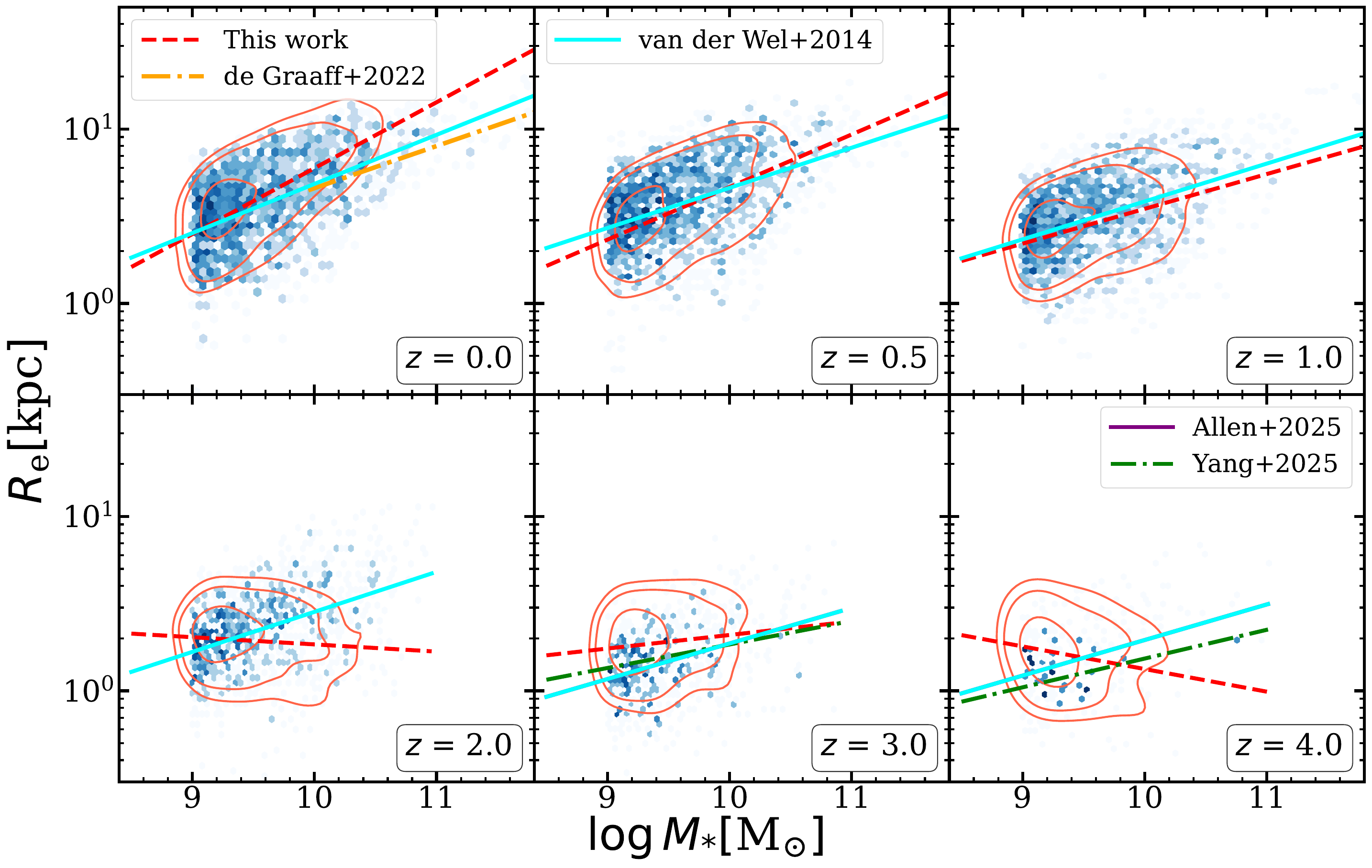}
    \caption{$M_{\ast}$--$R_{\mathrm{e}}$ distribution for the TNG50-SFGs with original (red contours) and corrected (blue colour-coded hexagonal bins) $R_{\mathrm{e}}$. The red dashed lines represent the best-fitting relations given in the form of Eq.~\ref{eq:eqRe} to the data points of red contours, while the cyan lines refer to the $M_{\ast}$--$R_{\mathrm{e}}$ relations from \protect\citet{vanderWel2014} ($0 \leq z \leq 2$) and \citet{Allen2025} ($3 \leq z \leq 4$). The orange and green lines show the $M_{\ast}$--$R_{\mathrm{e}}$ relations from the mock catalogue of \citet{deGraaff2022} ($z=0$) and from \citet{Yang2025} ($3 \leq z \leq 4$).}
    \label{fig:figA2}
\end{figure}

We follow the procedure of \citet{ManQiao2024} (their Sect.~3.2), who applied it to EAGLE-SFGs at $0.1 \leq z \leq 2.5$, and refer the reader to that work for details. In brief, at each redshift we fit the simulated $M_{\ast}$--$X$ distribution with the same functional form as the observed relation, compute the offset $\Delta X(M_{\ast}, z) = X^{\mathrm{fit}}_{\mathrm{sim}} - X_{\mathrm{obs}}$, and assign $X_{\mathrm{corr}} = X - \Delta X$ to each galaxy. When the observed relations are reported at redshifts different from ours, we interpolate them in redshift. Since our analysis extends to $z=4$, we complement the relations used by \citet{ManQiao2024} with JWST- and ALMA-based constraints at $z \geq 3$. The adopted relations are summarised in Table~\ref{tab:corrections}, and the original and corrected distributions are shown in Fig.~\ref{fig:figA2} and Figs.~\ref{fig:figA3}--\ref{fig:figA6}. These corrections modify the inputs of the observed relation, not the simulated galaxies: the radiative transfer is not recomputed, and the physical coupling between metallicity, dust content, SFR, and the UV and IR luminosities is not propagated. They therefore measure how sensitive the predicted IRX is to each property, not which physical process causes the discrepancy; the latter is addressed in Sects.~\ref{sec:indices} and \ref{sec:Physical_origins}.

By correcting each property individually while keeping the others unchanged, we test whether bringing that property into agreement with observations reduces the offset from the universal IRX relation.

\begin{table}
\caption{Observed scaling relations used to correct the TNG50-SFGs.}
\label{tab:corrections}
\centering
\footnotesize
\setlength{\tabcolsep}{2pt}
\begin{tabular*}{\columnwidth}{@{\extracolsep{\fill}}lll@{}}
\hline\hline
Quantity & $0 \leq z \leq 2$ & $3 \leq z \leq 4$ \\
\hline
$R_{\mathrm{e}}$ & \citet{vanderWel2014} & \citet{Allen2025} \\
$Z$ & \citet{Sanders2023} & \citet{Sanders2023} ($z=3$) \\
& & \citet{Stanton2026} ($z=4$) \\
SFR & \citet{Popesso2023} & \citet{Popesso2023} \\
IRX & \citet{Qin2019} ($z=0$) & \citet{Bouwens2020} \\
& \citet{Whitaker2014} ($z \geq 0.5$) & \\
$L_{\mathrm{IR}}$ & from SFR and IRX & from SFR and IRX \\
\hline
\end{tabular*}
\tablefoot{At $z=0$, we additionally test a direct calibration to the \textsc{DustPedia} $M_{\ast}$--$L_{\mathrm{IR}}$ relation \citep{Nersesian2019}.}
\end{table}

We first consider the observed $M_{\ast}$--$R_{\mathrm{e}}$ relation, which we parametrise following \citet{Shen2003}:

\begin{equation}
R_{\mathrm{e}}/{\mathrm{kpc}} = A \cdot {\left(\frac{M_\ast}{5 \times 10^{10}\,{\mathrm {M_\odot}}}\right)}^{\alpha_{R}},
\label{eq:eqRe}
\end{equation}

\noindent with $A$ and $\alpha_{R}$ taken from the late-type galaxies of \citet[][3D-HST/CANDELS; their Table~1]{vanderWel2014} at $0 \leq z \leq 2$, and from the SFGs of \citet[][CEERS, PRIMER-UDS and PRIMER-COSMOS, F200W; their Table~A1]{Allen2025} at $3 \leq z \leq 4$.

At $z \leq 1$, the sizes of the TNG50-SFGs are broadly consistent with observations, with massive galaxies being only slightly larger (Fig.~\ref{fig:figA2}). At $z \geq 2$, the simulated relation flattens and eventually develops a negative slope, with a scatter that increases with redshift and stellar mass, whereas the observed relation keeps a positive slope at all redshifts. A similar trend was reported for EAGLE-SFGs by \citet{ManQiao2024}. At the highest redshifts, repeated cycles of compaction and expansion in massive galaxies may contribute to the larger scatter and to the presence of very compact objects at the high-mass end \citep{Dekel2014,Lapiner2024, Cataldi2026}.

For gas-phase metallicity, we adopt at $0 \leq z \leq 3$ the $M_{\ast}$--$Z$ relation of \citet[][MOSDEF and SDSS]{Sanders2023}, using the same functional form and parameters as \citet[][their Eq.~2]{ManQiao2024}. At $z=4$, we use the JWST/EXCELS relation of \citet{Stanton2026}:

\begin{equation}
12 + \log {\mathrm {(O/H)}} = \beta_{R} \, \log (M_{\ast}/10^9\, \mathrm{M_{\odot}}) + Z_9,
\label{eq:eqZ2}
\end{equation}

\noindent interpolating their best-fitting $\beta_{R}$ and $Z_9$ between $z=3.2$ and $z=5.5$.

TNG50-SFGs are systematically more metal-rich than observed SFGs at all redshifts, by $\sim$0.2--0.3\, dex (Fig.~\ref{fig:figA3}), although the offset is smaller than for EAGLE-SFGs, which reach $\sim$0.5\,dex at $z\approx0$ \citep[see their Fig.~3]{ManQiao2024}. While in EAGLE the offset decreases with increasing redshift, in TNG50 it remains approximately constant, decreasing only at the high-mass end at $z=4$. This offset indicates a tension with observations, although observed gas-phase metallicities carry systematic uncertainties of up to $\sim$0.7\,dex at comparable redshifts, driven by sample selection, aperture effects and, above all, the choice of metallicity calibrator \citep{Genzel2015, DeRossi2017, Blanc2019, Sanders2025}.

For the correction to $L_{\mathrm{IR}}$, we combine the corrections to the SFR and IRX. For the SFR, we adopt the $M_{\ast}$--SFR relation of \citet{Popesso2023}, which is calibrated over $0 < z < 6$ and therefore covers our full redshift range. As found for EAGLE \citep{ManQiao2024}, the TNG50-SFGs have systematically lower SFRs than observed SFGs over most of the mass range, except at the highest masses at $z\leq1$ (Fig.~\ref{fig:figA4}). Although TNG reproduces the observed slope of the star-forming main sequence, its normalisation at $10^{9}$--$10^{10.5}\,\mathrm{M_{\odot}}$ is lower than observed by at least 0.2~dex and up to 0.5~dex \citep{Donnari2019}. At the same time, SFRs derived from combined UV+IR luminosities \citep[e.g.][]{Whitaker2014,Popesso2023} may overestimate the normalisation \citep{Katsianis2020, Leja2022}, so part of this offset may also reflect observational systematics, in addition to limitations of the simulation.

For the remaining quantities entering the IRX analysis, we adopt the $M_{\ast}$--IRX relations of \citet{Qin2019} at $z=0$ and \citet{Whitaker2014} at $0.5 \leq z \leq 2$, both described by a second-order polynomial in $\log M_{\ast}$ \citep[][their Eq.~6]{ManQiao2024}. At $z \geq 3$ we use the ALMA/ASPECS relation of \citet{Bouwens2020},

\begin{equation}
\log {\mathrm{IRX}} = a \, \log (M_{\ast}/M_s),
\label{eq:eqIRX2}
\end{equation}

\noindent with $a = 0.97$ and $\log(M_s/\mathrm{M_{\odot}}) = 9.15$. Similar to EAGLE-SFGs, the TNG50-SFGs have systematically lower IRX than observed at $z=0$, by $\sim$0.3--0.4\,dex. At $z \geq 0.5$, the offset becomes mass dependent, with an excess of IRX at the low-mass end and a deficit at the high-mass end, of up to $\sim$0.5\,dex in either direction (Fig.~\ref{fig:figA5}).

From the relation ${\mathrm{SFR}} = k\,(L_{\mathrm{IR}} + 2.2\,L_{\mathrm{UV}})$, with $k=1.09 \times 10^{-10}\,\mathrm{M_\odot\,yr^{-1}\,L_\odot^{-1}}$ \citep{Bell2005}, it follows that $L_\mathrm{IR}=(\mathrm{SFR}/k)\,\mathrm{IRX}/(\mathrm{IRX}+2.2)$. We propagate the corrections to SFR and IRX differentially: the corrected $L_\mathrm{IR}$ of each galaxy is its simulated $L_\mathrm{IR}$ multiplied by the ratio of this expression evaluated with the corrected and with the original SFR and IRX, and the corrected $L_\mathrm{UV}$ follows from $L_\mathrm{IR}/\mathrm{IRX}$. With this prescription, $L_\mathrm{IR}$ remains equal to the simulated value when no correction is required. At $z=0$, where direct observational constraints are available, we also compare with the median $M_{\ast}$--$L_{\mathrm{IR}}$ relation of \textsc{DustPedia} SFGs \citep{Nersesian2019}. The \textsc{DustPedia} relation lies above that of the TNG50-SFGs, with an offset in $L_{\mathrm{IR}}$ that increases with stellar mass from $\sim$0.3 to $\sim$0.5\,dex (Fig.~\ref{fig:figA6}). As an alternative calibration at $z=0$, we therefore shift the simulated $L_{\mathrm{IR}}$ directly onto the \textsc{DustPedia} median relation, and use it to test the robustness of the results in Sect.~\ref{sec:IRX_corrected}.

\subsection{IRX relation with corrected parameters}
\label{sec:IRX_corrected}

We now re-examine the predicted universal IRX relation, correcting one parameter at a time: $R_{\mathrm{e}}$, $Z$, and $L_{\mathrm{IR}}$ together with IRX. This allows us to assess which of these parameters the predicted relation is most sensitive to, and how much each contributes to the discrepancy with the observed universal IRX relation.

\begin{figure}
    \centering
    \includegraphics[width=\columnwidth]{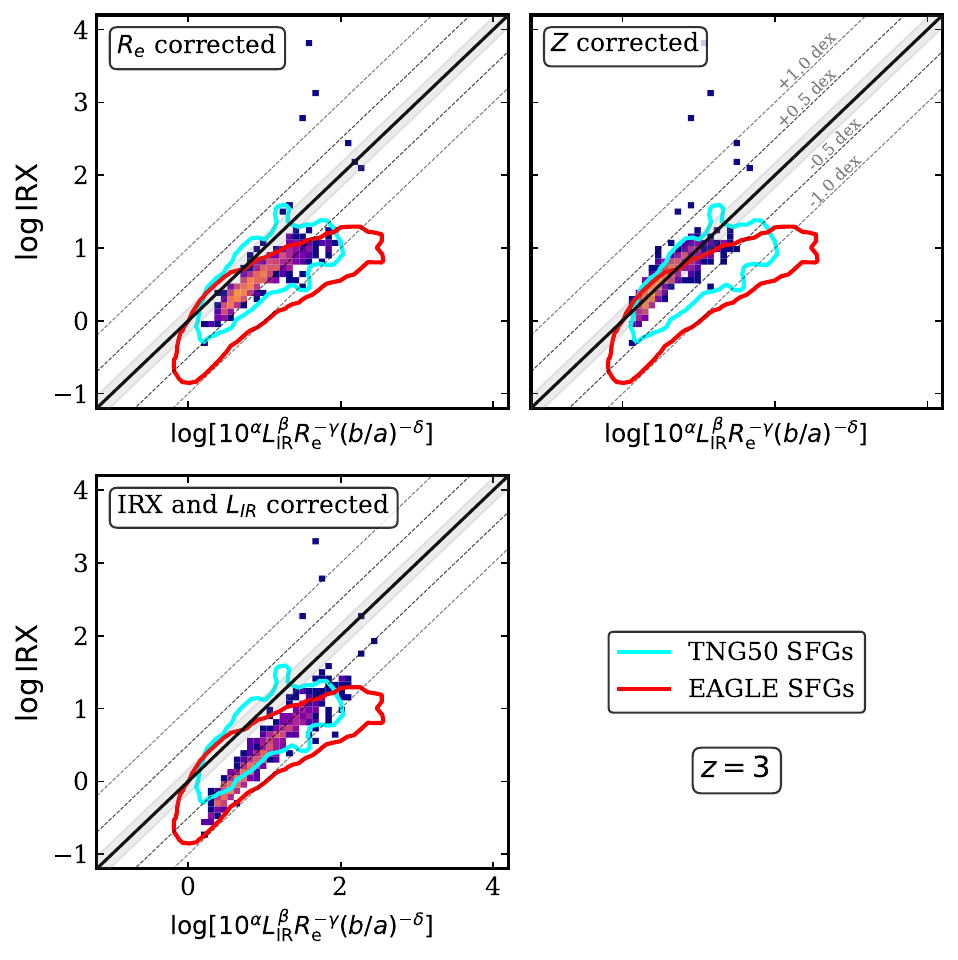}
    \caption{Predicted universal IRX relation after applying corrections to the half-light radius ($R_\mathrm{e}$; top-left panel), metallicity ($Z$; top-right panel), and both $L_\mathrm{IR}$ and IRX (bottom-left panel). For clarity, we show only the case at $z=3$, where the deviations at the high-IRX end are largest. The cyan and red contours indicate the distributions obtained using the original parameters for our sample and the EAGLE-SFGs, respectively.}
    \label{fig:fig3}
\end{figure}

Figure~\ref{fig:fig3} shows the results at $z=3$, where the deviations from the observed relation are largest at the high-IRX end; all redshifts are shown in Fig.~\ref{fig:figA7}. The metallicity correction produces the largest improvement, bringing the simulated galaxies into much closer agreement with the universal IRX relation of \citet{Qin2019} at $z\geq2$, while it leaves the offset at $z=0$ largely unchanged. This partly reflects the structure of the relation itself: metallicity enters all four power-law indices ($\alpha$, $\beta$, $\gamma$, and $\delta$), whereas $R_{\mathrm{e}}$ and $L_{\mathrm{IR}}$ each enter a single term, so the predicted IRX is most sensitive to the simulated gas-phase metallicity. This identifies the metallicity offset as the largest contributor to the difference between the IRX measured from the simulations and the IRX predicted by the universal relation under our calibration scheme, but does not by itself establish metallicity as its main physical origin. We return to this point in Sect.~\ref{sec:calibration}, where we show that a uniform shift of $-0.2$~dex in the TNG50 metallicities is enough to bring the simulated galaxies onto the universal IRX relation at $z\geq2$, but not at $z=0$.

In contrast, correcting $R_\mathrm{e}$ to the relations of \citet{vanderWel2014} and \citet{Allen2025} has the smallest impact, because the simulated sizes are close to these relations at $z \leq 1$. This does not mean that sizes play no role: relative to the calibration sample of \citet{Qin2019}, whose sizes are measured differently, the TNG50-SFGs are larger at fixed stellar mass and $L_\mathrm{IR}$, and the size term contributes significantly to the offset at $z\approx0$ (Sect.~\ref{sec:z0}). The combined correction to $L_\mathrm{IR}$ and IRX produces shifts in the predicted IRX, but does not systematically improve the agreement with the observed relation and can even lead to larger deviations at some redshifts. At $z=0$, calibrating $L_{\mathrm{IR}}$ directly to \textsc{DustPedia} yields consistent results (Fig.~\ref{fig:figA7}). We note that the corrections to $L_{\mathrm{IR}}$ and IRX rely on empirical scaling relations that may carry substantial uncertainties, particularly in the SFR normalisation (Sect.~\ref{sec:corrections}). Overall, although the corrections to $R_\mathrm{e}$ and $L_{\mathrm{IR}}$+IRX produce noticeable shifts in the individual scaling relations (Figs.~\ref{fig:figA2} and \ref{fig:figA5}), neither accounts for the bulk of the discrepancy with the observed universal IRX relation.

\subsection{Metallicity power-law indices}
\label{sec:indices}

\begin{figure}
    \centering
    \includegraphics[width=0.8\columnwidth]{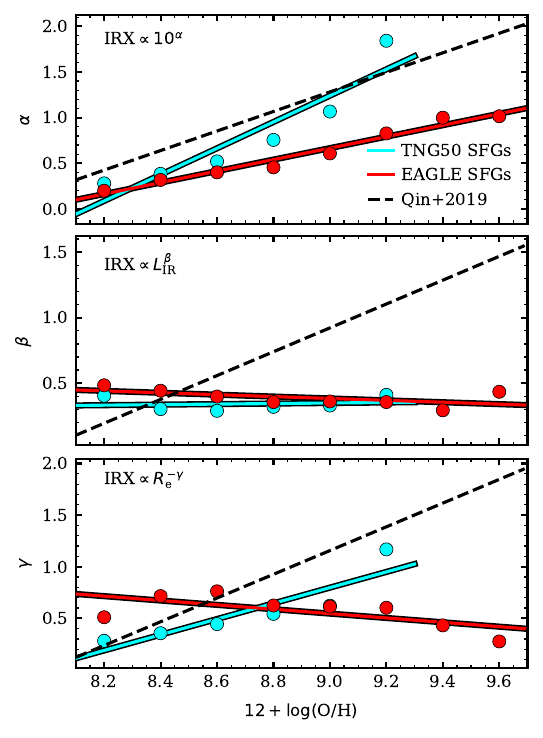}
    \caption{Power-law indices from the predicted universal IRX relation, $\alpha$, $\beta$, and $\gamma$ as a function of gas-phase metallicity for the sample TNG50-SFGs with the original galaxy parameters. The cyan solid circles represent the best-fitting values, while the red solid circles represent the best-fitting values for the EAGLE-SFGs of \citet{ManQiao2024}, obtained in the same way. The cyan and red solid lines denote the corresponding linear fits (for TNG50-SFGs, Eq.~\ref{eq:indices_tng}). The dashed black lines are from the results reported in \citet{Qin2019}. In all cases, the inclination index is $\delta(Z)$ of \citet{Qin2019} at $b/a=0.6$.}
    \label{fig:fig4}
\end{figure}

Given that metallicity plays a key role in determining the power-law indices (see Eq.~\ref{eq:eqIndex}), we examine the deviation between the indices reported by \citet{Qin2019} and those derived from TNG50-SFGs. To this end, the sample (all redshifts combined) is divided into six bins of $0.2$~dex spanning $8.1 < 12 + \log (\mathrm{O/H}) < 9.3$, and the universal IRX relation (Eq.~\ref{eq:eq3} and Eq.~\ref{eq:eqIndex}) is fitted independently within each bin. Since all galaxies are assigned the same axis ratio, $b/a = 0.6$, the inclination term cannot be fitted independently of the normalisation; we therefore adopt the metallicity-dependent $\delta(Z)$ of \citet{Qin2019} (Eq.~\ref{eq:eqIndex}) for each galaxy, so that $\alpha$ is directly comparable with the observed normalisation. The resulting best-fitting indices are shown in Fig.~\ref{fig:fig4} as cyan symbols, together with their linear fits (cyan lines). For comparison, we include the observational constraints from \citet{Qin2019} (black dashed lines) and the results for EAGLE-SFGs from the catalogue of \citet{ManQiao2024}, fitted in the same way (red symbols and solid lines).

The linear fits define the empirical IRX relation of the TNG50-SFGs, with power-law indices

\begin{equation}
\begin{aligned}
\alpha_\mathrm{TNG} &= 1.44\,\log (Z/Z_\odot) + 0.79,\\
\beta_\mathrm{TNG} &= 0.02\,\log (Z/Z_\odot) + 0.34,\\
\gamma_\mathrm{TNG} &= 0.76\,\log (Z/Z_\odot) + 0.56,
\end{aligned}
\label{eq:indices_tng}
\end{equation}

\noindent to be compared with $\alpha=1.07\,\log (Z/Z_\odot) +0.95$, $\beta=0.91\,\log (Z/Z_\odot)+0.64$ and $\gamma=1.15\,\log (Z/Z_\odot) +0.80$ in \citet{Qin2019} (Eq.~\ref{eq:eqIndex}).

The three indices differ in distinct ways. The normalisation $\alpha_\mathrm{TNG}$ is lower than the observed one at sub-solar metallicity and increases more steeply with metallicity ($1.44$ versus $1.07$ per dex), reaching the observed value at $12+\log(\mathrm{O/H})\approx9.1$. The $L_\mathrm{IR}$ slope is $\beta_\mathrm{TNG}\approx0.34$, about half of the observed value at solar metallicity, and shows no dependence on metallicity, whereas it decreases slightly with metallicity in EAGLE-SFGs. The size slope $\gamma_\mathrm{TNG}$ increases with metallicity, as observed, but with a lower normalisation and a shallower dependence ($0.76$ versus $1.15$ per dex). Fitting the indices separately at each redshift (not shown) gives the same behaviour, with $\beta_\mathrm{TNG}=0.2$--$0.4$ in all metallicity bins and at all redshifts, so these trends are not an artefact of combining galaxies from different epochs. Fitting $\alpha_\mathrm{TNG}$, $\beta_\mathrm{TNG}$ and $\gamma_\mathrm{TNG}$ jointly to all galaxies, without binning in metallicity, gives consistent indices, with $\beta_\mathrm{TNG}=0.34-0.02\,\log(Z/Z_\odot)$ and $\alpha_\mathrm{TNG}$ and $\gamma_\mathrm{TNG}$ within $\approx0.1$ of Eq.~\ref{eq:indices_tng} at solar metallicity, so the metallicity independence of $\beta_\mathrm{TNG}$ does not depend on the fitting method.

In the observed relation, $\beta$ and $\gamma$ increase with metallicity in nearly fixed proportion, so that $\gamma/\beta\simeq1.25$ at all metallicities (Eq.~\ref{eq:eqIndex}) and dust obscuration is regulated by the single combination $L_\mathrm{IR}/R_\mathrm{e}^{1.25}$. In TNG50-SFGs, $\beta_\mathrm{TNG}$ is constant while $\gamma_\mathrm{TNG}$ increases with metallicity, so $\gamma_\mathrm{TNG}/\beta_\mathrm{TNG}$ rises from $\approx1.0$ at $12+\log(\mathrm{O/H})=8.4$ to $\approx1.65$ at solar metallicity and $\approx2.7$ at $12+\log(\mathrm{O/H})=9.2$; no single combination of $L_\mathrm{IR}$ and $R_\mathrm{e}$ describes the simulated galaxies at all metallicities. For EAGLE-SFGs, \citet{ManQiao2024} find $\gamma/\beta\sim1.5$. We discuss the physical origin of these differences in Sect.~\ref{sec:interpretation}.

These results suggest that the weak dependence of IRX on $L_\mathrm{IR}$ ($\beta$) plays a more significant role than the dependence on $R_\mathrm{e}$ ($\gamma$) in driving the mismatch between simulations and observations; we quantify the contribution of each index in Sect.~\ref{sec:budget}.

\subsection{Decomposition of the IRX offsets}
\label{sec:Physical_origins}

The offsets in Fig.~\ref{fig:fig2} change sign with redshift, from $\Delta\log\mathrm{IRX}\approx+0.5$~dex at $z=0$ to $\approx-0.25$~dex at $z\geq2$. Because both the observed relation (Eq.~\ref{eq:eq3}) and the TNG50 relation (Sect.~\ref{sec:indices}, Eq.~\ref{eq:indices_tng}) are linear in the logarithm of their inputs, this offset can be decomposed exactly into the contributions of the individual indices. We write both relations as

\begin{equation}
\log\mathrm{IRX}_\mathrm{model} = \mathcal{A}(Z) + \beta(Z)\,\ell - \gamma(Z)\,r,
\label{eq:irx_log}
\end{equation}

\noindent where $\ell\equiv\log(L_\mathrm{IR}/10^{10}\,\mathrm{L_\odot})$, $r\equiv\log(R_\mathrm{e}/\mathrm{kpc})$, and $\mathcal{A}\equiv\alpha-\delta\log(b/a)$ is the normalisation at the pivot adopted by \citet{Qin2019}, $L_\mathrm{IR}=10^{10}\,\mathrm{L_\odot}$ and $R_\mathrm{e}=1$~kpc, evaluated at $b/a=0.6$. Replacing each index of the observed relation (subscript Q) by the corresponding index of the TNG50 relation (subscript TNG; Eq.~\ref{eq:indices_tng}) gives

\begin{equation}
\Delta\log\mathrm{IRX} = \Delta_\mathrm{TNG} + \Delta_\alpha + \Delta_\beta + \Delta_\gamma,
\label{eq:budget}
\end{equation}

\noindent with

\begin{equation}
\begin{aligned}
\Delta_\alpha &= \mathcal{A}_\mathrm{TNG}(Z) - \mathcal{A}_\mathrm{Q}(Z),\\
\Delta_\beta &= \left[\beta_\mathrm{TNG}(Z) - \beta_\mathrm{Q}(Z)\right]\ell,\\
\Delta_\gamma &= \left[\gamma_\mathrm{Q}(Z) - \gamma_\mathrm{TNG}(Z)\right] r,
\end{aligned}
\label{eq:budget_terms}
\end{equation}

\noindent and $\Delta_\mathrm{TNG}$ the residual of each galaxy about the TNG50 relation itself. Eq.~\ref{eq:budget} holds exactly for every galaxy. Each of $\Delta_\beta$ and $\Delta_\gamma$ is the product of an index difference and the distance of the galaxy from the pivot in $L_\mathrm{IR}$ or $R_\mathrm{e}$. A difference in slope therefore produces no offset for galaxies at the pivot, but an offset that grows linearly with their distance from it, with opposite signs on either side.

The split of $\Delta\log\mathrm{IRX}$ between $\Delta_\alpha$, $\Delta_\beta$ and $\Delta_\gamma$ depends on the choice of pivot, for which we follow the convention of \citet{Qin2019}; the total offset and $\Delta_\mathrm{TNG}$ are independent of this choice.

\subsubsection{Index budget}
\label{sec:budget}

\begin{figure}
\resizebox{\hsize}{!}{\includegraphics{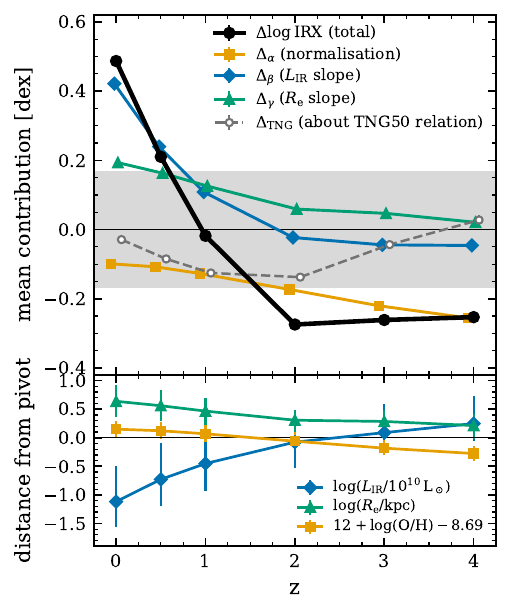}}
\caption{Index budget of the IRX residuals of the TNG50-SFGs with respect to the universal IRX relation. \textit{Top:} mean $\Delta\log\mathrm{IRX}$ (black) and its decomposition into the four terms of Eq.~\ref{eq:budget}, which add up to the total at each redshift. The bootstrap uncertainties on the means (16th--84th percentiles) are smaller than the symbols; the shaded band marks the $\pm0.17$~dex scatter of the observed relation. \textit{Bottom:} median distance of the TNG50-SFGs from the pivot of the relation in $L_\mathrm{IR}$ and $R_\mathrm{e}$, and their median metallicity relative to solar, with 16th--84th percentile ranges.}
\label{fig:budget}
\end{figure}

Figure~\ref{fig:budget} shows the mean of each term of Eq.~\ref{eq:budget} as a function of redshift; we use means rather than medians because only means are additive. The mean residual about the TNG50 relation itself is small at all redshifts ($|\langle\Delta_\mathrm{TNG}\rangle|\leq0.14$~dex), with a scatter of 0.21--0.35~dex, larger than the 0.17~dex of the observed relation. The TNG50-SFGs thus follow a single relation of the same functional form over $0\leq z\leq4$, and the offsets in Fig.~\ref{fig:fig2} arise from the differences between its indices and those of \citet{Qin2019}, combined with the location of the simulated galaxies relative to the pivot.

At $z=0$, the offset of $+0.49$~dex is dominated by the $L_\mathrm{IR}$ term ($\Delta_\beta=+0.42$~dex), with a smaller contribution from the size term ($\Delta_\gamma=+0.19$~dex) partly compensated by the normalisation ($\Delta_\alpha=-0.10$~dex). The TNG50 slope $\beta_\mathrm{TNG}\approx0.34$ is independent of metallicity, whereas $\beta_\mathrm{Q}\approx0.78$ at the median metallicity of the $z=0$ sample ($12+\log(\mathrm{O/H})=8.84$). Since the median TNG50-SFG at $z=0$ lies $1.1$~dex below the pivot ($\log L_\mathrm{IR}=8.88$, see Table~\ref{tab:sample_properties}), the steeper observed slope predicts a much lower IRX than the simulated galaxies have, and $\Delta\log\mathrm{IRX}>0$. The size term has the same origin: $\gamma_\mathrm{TNG}<\gamma_\mathrm{Q}$ and the TNG50-SFGs are larger than the 1~kpc pivot (median $\log R_\mathrm{e}=0.64$), so the observed relation predicts a stronger decrease of IRX with size than the simulated galaxies show.

Towards higher redshift, the TNG50-SFGs become more IR-luminous and more compact. The median $\log L_\mathrm{IR}$ increases from $8.88$ at $z=0$ to $9.92$ at $z=2$ and $10.25$ at $z=4$, close to the pivot, so the slope terms vanish ($|\Delta_\beta|,\,|\Delta_\gamma|\lesssim0.06$~dex at $z\geq2$). At $z\approx1$ the remaining positive slope terms ($\Delta_\beta + \Delta_\gamma=+0.24$~dex) are cancelled by the negative normalisation and residual terms, which explains the apparent agreement with the observed relation at intermediate redshift (Fig.~\ref{fig:fig2}). At $z\geq2$ the offset is dominated by the normalisation, with $\Delta_\alpha=-0.17$, $-0.22$ and $-0.26$~dex at $z=2$, 3 and 4; at $z=2$, the residual about the TNG50 relation contributes a further $\approx-0.14$~dex. The normalisation term depends on metallicity alone: from the fitted indices, $\Delta_\alpha\simeq-0.16+0.37\,[12+ \log(\mathrm{O/H}) - 8.69]$, which vanishes at $12 + \log(\mathrm{O/H}) \approx 9.1$. Because the normalisation of the TNG50 relation increases more steeply with metallicity than the observed one ($1.44$ versus $1.07$ per dex; Fig.~\ref{fig:fig4}), $\Delta_\alpha$ becomes increasingly negative as the median metallicity of the TNG50-SFGs decreases from $8.84$ at $z=0$ to $8.41$ at $z=4$. In other words, at fixed $L_\mathrm{IR}$ and $R_\mathrm{e}$, sub-solar TNG50-SFGs are less obscured than the observed relation predicts for their metallicity, an offset equivalent to a shift of $\approx0.2$~dex in the metallicity scale (Sect.~\ref{sec:calibration}).

The redshift dependence of $\Delta\log\mathrm{IRX}$ in Fig.~\ref{fig:fig2} is therefore not primarily an evolution of dust attenuation in TNG50, but the combination of two effects that dominate in different regimes. At low redshift, the flat $L_\mathrm{IR}$ dependence of TNG50 is amplified by the large distance of the simulated galaxies from the pivot, which lies close to the lower limit of the $L_\mathrm{IR}$ range in which the relation was calibrated. At high redshift, the galaxies sit near the pivot, and the metallicity dependence of the normalisation dominates.

\subsubsection{Calibration range of the relation}
\label{sec:calibration}

\begin{figure}
\resizebox{\hsize}{!}{\includegraphics{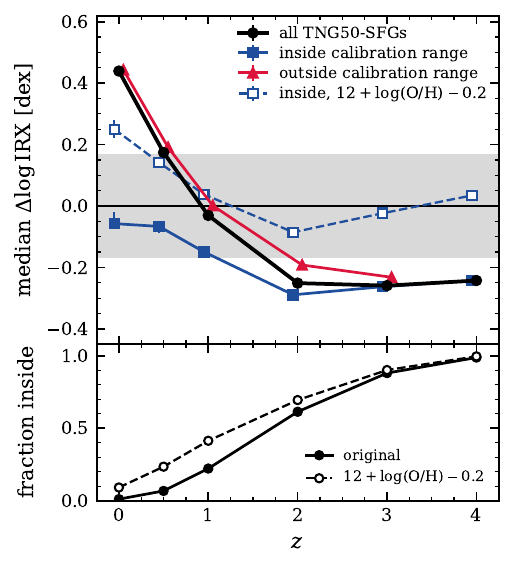}}
\caption{\textit{Top:} median IRX residual of the TNG50-SFGs as a function of redshift, for the full sample and for galaxies inside and outside the range in which the universal relation was calibrated ($\log(L_\mathrm{IR}/\mathrm{L_\odot})\geq9.63$ and $12+\log(\mathrm{O/H})\leq8.83$; Sect.~\ref{sec:calibration}), with the original and the shifted ($-0.2$~dex) metallicities. The bootstrap uncertainties on the medians (16th--84th percentiles) are mostly smaller than the symbols; the shaded band marks the $\pm0.17$~dex scatter of the observed relation. \textit{Bottom:} fraction of TNG50-SFGs inside the calibration range.}
\label{fig:calibration}
\end{figure}

The universal IRX relation was calibrated on $\sim$32,000 local SFGs ($0.04 < z < 0.15$), of which 97.5\% have $\log(L_\mathrm{IR}/\mathrm{L_\odot}) \geq 9.63$ and 97.5\% have $12+\log(\mathrm{O/H}) \leq 8.83$, the latter close to the saturation of the N2 metallicity indicator. Applying the relation to galaxies outside this range requires extrapolating it. Fig.~\ref{fig:calibration} shows the median $\Delta\log\mathrm{IRX}$ of the TNG50-SFGs separately for galaxies inside and outside this range. The fraction of TNG50-SFGs inside the calibration range increases from 1\% at $z=0$ to 22\% at $z=1$, 61\% at $z=2$, and 99\% at $z=4$: at low redshift, the simulated galaxies are predominantly fainter in $L_\mathrm{IR}$ and more metal-rich than the calibration sample.

Galaxies inside the calibration range never lie above the relation by more than its scatter: their median offset decreases smoothly from $\approx0$ at $z\leq0.5$ ($-0.06$~dex at $z=0$ and $-0.07$~dex at $z=0.5$) to $-0.15$~dex at $z=1$ and $\approx-0.25$~dex at $z\geq2$. The positive offsets at $z\leq0.5$ arise entirely from galaxies outside the calibration range ($+0.44$~dex at $z=0$ and $+0.19$~dex at $z=0.5$), and as the fraction of galaxies inside increases with redshift, the median offset of the full sample converges to that of the galaxies inside. This is consistent with the index budget of Sect.~\ref{sec:budget}: the slope terms that dominate at low redshift grow with the distance from the pivot, which for most TNG50-SFGs at $z\lesssim0.5$ lies outside the range probed by the calibration sample.

The negative offset at high redshift, in contrast, is found inside the calibration range and is driven by the normalisation term, which depends on metallicity alone. Lowering the TNG50 metallicities by 0.2~dex, well within the systematic uncertainties of observed metallicity calibrations (Sect.~\ref{sec:corrections}), brings the galaxies inside the calibration range onto the relation at $z\geq2$ ($-0.09$, $-0.02$ and $+0.03$~dex at $z=2$, 3 and 4). The same shift, however, increases the offset of the galaxies inside the range at $z=0$ to $+0.25$~dex. No single metallicity offset therefore reconciles both regimes: the high-redshift deficit is consistent with a difference in metallicity scale, whereas the low-redshift excess is not.

\subsubsection{Comparison with the calibration sample at $z\approx0$}
\label{sec:z0}

Sections~\ref{sec:budget} and \ref{sec:calibration} compare the TNG50 relation with the observed one; at $z\approx0$, the simulated galaxies can also be compared directly with the galaxies used to calibrate it. We use the SDSS-WISE-GALEX sample of \citet{Qin2019} and include the EAGLE-SFGs of \citet{ManQiao2024} at $z=0.1$, restricting all samples to $M_\star>10^9\,\mathrm{M_\odot}$, the stellar-mass limit of the TNG50 sample. From Eq.~\ref{eq:irx_log}, the residual of each galaxy is $\Delta\log\mathrm{IRX}=\log\mathrm{IRX}-\mathcal{A}(Z)-\beta(Z)\,\ell+\gamma(Z)\,r$, with the measured axis ratios for the observed galaxies and $b/a=0.6$ for the simulations. Writing $D[x]\equiv\langle x\rangle_\mathrm{sim}-\langle x\rangle_\mathrm{obs}$ for the difference between the mean values of a quantity in simulated and observed galaxies with the same stellar mass,

\begin{equation}
D[\Delta\log\mathrm{IRX}] = D[\log\mathrm{IRX}] - D[\mathcal{A}] - D[\beta\,\ell] + D[\gamma\,r],
\label{eq:z0_terms}
\end{equation}

\noindent so that the offset between simulated and observed galaxies separates exactly into the contributions of their IRX, their normalisation (set by metallicity and inclination), their $L_\mathrm{IR}$ and their size.

\begin{figure}
\resizebox{\hsize}{!}{\includegraphics{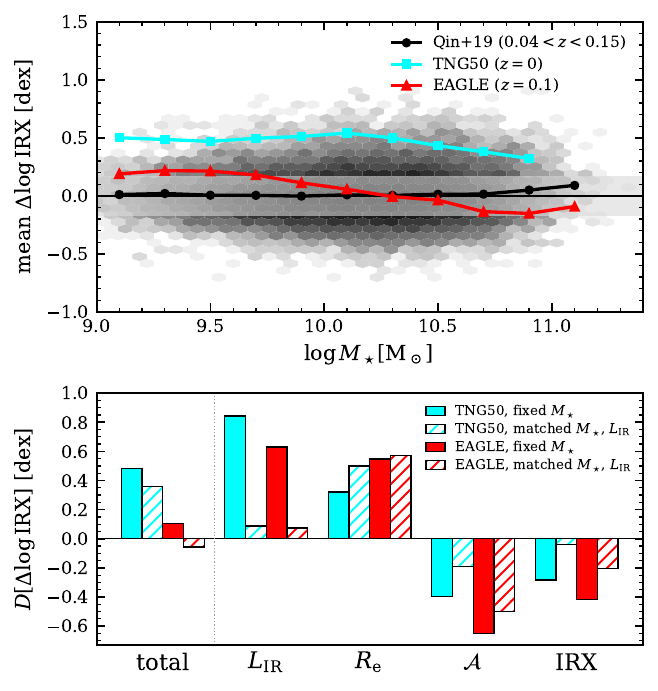}}
\caption{\textit{Top:} mean IRX residual as a function of stellar mass for the SDSS-WISE-GALEX calibration sample of \citet{Qin2019} (black; grey map shows its distribution), TNG50-SFGs at $z=0$ (cyan) and EAGLE-SFGs at $z=0.1$ (red). All samples are restricted to $M_\star>10^9\,\mathrm{M_\odot}$, the stellar-mass limit of the TNG50 sample. The shaded band marks the $\pm0.17$~dex scatter of the relation. \textit{Bottom:} decomposition of the difference between the mean residuals of simulated and observed galaxies (Eq.~\ref{eq:z0_terms}) into the contributions of $L_\mathrm{IR}$, size ($R_\mathrm{e}$), normalisation ($\mathcal{A}$, including the inclination term) and IRX, for TNG50 (cyan) and EAGLE (red), at fixed stellar mass (solid bars; bins of 0.2~dex) and at matched stellar mass and IR luminosity (hatched bars; cells of 0.25~dex in both). The contributions are weighted by the number of simulated galaxies in each bin or cell, and add up to the total (left).}
\label{fig:z0}
\end{figure}

The observed galaxies lie on the relation at all stellar masses, as expected for the calibration sample, whereas the TNG50-SFGs lie $\approx+0.5$~dex above it up to $M_\star\approx10^{10.4}\,\mathrm{M_\odot}$ and the EAGLE-SFGs between $+0.2$ and $-0.2$~dex (Fig.~\ref{fig:z0}, top). At fixed stellar mass, the offset of TNG50 ($+0.48$~dex on average) is dominated by the $L_\mathrm{IR}$ contribution ($+0.84$~dex): the TNG50-SFGs are $\approx1$~dex fainter in $L_\mathrm{IR}$ than observed galaxies of the same mass, partly because of the lower SFRs of TNG50 (Sect.~\ref{sec:corrections}) and partly because the calibration sample, which requires WISE and GALEX detections, is biased towards IR-bright galaxies. The size contribution adds $+0.32$~dex, while the higher metallicities of the simulated galaxies, through the normalisation ($-0.40$~dex), and their lower IRX ($-0.28$~dex) partly compensate. EAGLE shows the same structure with a smaller $L_\mathrm{IR}$ deficit, and a total offset of $+0.11$~dex ($+0.63$ from $L_\mathrm{IR}$, $+0.55$ from size, $-0.65$ from the normalisation and $-0.42$ from IRX).

Matching the simulated and observed galaxies in both stellar mass and $L_\mathrm{IR}$ removes the $L_\mathrm{IR}$ contribution (Fig.~\ref{fig:z0}, bottom, hatched bars). In this regime, which contains 7\% of the TNG50-SFGs (more than the 1\% inside the calibration range of Sect.~\ref{sec:calibration}, because the matching imposes no metallicity limit and includes the few observed galaxies below $\log L_\mathrm{IR}=9.63$) and 36\% of the EAGLE-SFGs, the IRX of TNG50 agrees with the observed one to within 0.04~dex; where a like-for-like comparison is possible, the dust attenuation of TNG50 is consistent with observations. The remaining offset of $+0.36$~dex arises from the size contribution ($+0.50$~dex): the TNG50-SFGs have effective radii 0.33~dex larger than observed galaxies of the same stellar mass and IR luminosity, and metallicities 0.22~dex higher, which increase $\gamma$. We note that the sizes are measured differently in the two samples: \citet{Qin2019} use $r$-band semi-major-axis half-light radii from bulge+disc decompositions \citep{Simard2011}, whereas for the TNG50-SFGs we derive half-light radii from the stellar half-mass radii using the light-to-mass size ratios of \citet{Suess2019a, Suess2019b} (Sect.~\ref{sec:parameters}). For EAGLE, the mean offset at matched $M_\star$ and $L_\mathrm{IR}$ is only $-0.06$~dex, but this agreement results from the following cancellation: the IRX is 0.20~dex lower than observed, which is compensated by sizes 0.24~dex larger and metallicities 0.42~dex higher.

These results also explain why varying the DTM ratio does not remove the offset (Sect.~\ref{sec:f_dust_var}): since $f_\mathrm{dust}$ changes $L_\mathrm{IR}$ by at most $\approx0.3$~dex at $z=0$, the TNG50-SFGs remain well below the calibration sample in $L_\mathrm{IR}$ for any $f_\mathrm{dust}$. Nor does a higher $f_\mathrm{dust}$ steepen the $L_\mathrm{IR}$ dependence: fitting the 100-galaxy subsample (not shown), $\beta_\mathrm{TNG}$ remains nearly independent of metallicity for all $f_\mathrm{dust}$, and its value at solar metallicity decreases by $\approx0.2$ between $f_\mathrm{dust}=0.1$ and 1, moving further away from the observed value. Where simulated and observed galaxies can be compared directly, the dust attenuation of TNG50 at $z=0$ is therefore consistent with observations; the offset results from comparing galaxies that are fainter in $L_\mathrm{IR}$ and more extended than the calibration sample with a relation whose $L_\mathrm{IR}$ and size dependences are steeper than those of the simulation. The flat $L_\mathrm{IR}$ dependence itself may, however, originate in the dust post-processing (Sect.~\ref{sec:interpretation}).

\section{Discussion}
\label{sect:discussion} 

\subsection{Physical interpretation of the TNG50 relation}
\label{sec:interpretation}

A key point of comparison between the universal IRX relation and earlier attempts to parametrise IRX \citep[e.g.][]{Desert1990, Dale2002, Chanial2007} lies in the physical quantity assumed to regulate dust obscuration. In those models, IRX is primarily linked to the IR luminosity surface density, $\Sigma_{\mathrm{IR}} = L_{\mathrm{IR}}/R_{\mathrm{e}}^{2}$. In contrast, the universal IRX relation implies a weaker dependence on galaxy size, with $\gamma/\beta \simeq 1.25$ at all metallicities \citep{Qin2019}. Accounting for potential systematic uncertainties in $L_{\mathrm{IR}}$ measurements may shift this value to $\gamma/\beta \sim 1.3$--$1.5$, but the deviation from a purely surface-density-driven scenario remains. In TNG50-SFGs, $\gamma_\mathrm{TNG}/\beta_\mathrm{TNG}$ is not constant but increases from $\approx1.0$ to $\approx2.7$ with metallicity (Sect.~\ref{sec:indices}), because $\beta_\mathrm{TNG}$ is independent of metallicity while $\gamma_\mathrm{TNG}$ is not.

A further complication arises from the choice of characteristic radius. While $R_{\mathrm{e}}$ traces the stellar distribution, the IR emission is more closely associated with regions of active star formation and with the distribution of dust \citep{Lyu2026}. Ideally, one should therefore consider the half-IR radius, $R_{\mathrm{IR}}$, rather than $R_{\mathrm{e}}$. From a theoretical standpoint, the IRX can be related to the dust surface density, $\Sigma_{\mathrm{dust}}$, which in turn depends on both metallicity and star-forming gas content. Given that the dust-to-gas ratio scales with metallicity (Eq.~\ref{eq:eq1}), one expects $\Sigma_{\mathrm{dust}} \propto (Z/Z_{\odot}) \times \Sigma_{\mathrm{gas}}$. Combining this with the Kennicutt--Schmidt relation, $\Sigma_{\mathrm{SFR}} \propto \Sigma_{\mathrm{gas}}^{n}$, leads to

\begin{equation}
\mathrm{IRX} \propto \Sigma_{\mathrm{dust}}\propto \left(\frac{Z}{Z_{\odot}}\right)\times\left(\frac{\mathrm{SFR}}{R_{\mathrm{IR}}^2}\right)^{1/n},
\label{eq:eq9}
\end{equation}

where $n \approx 1.42$ (i.e. $1/n \approx 0.7$). For comparison, the universal IRX relation (Eqs.~\ref{eq:eq3} and \ref{eq:eqIndex}) can be recast in a similar form,

\begin{equation}
\mathrm{IRX} \propto \left(\frac{Z}{Z_{\odot}}\right)^{1.07}\times\left(\frac{\mathrm{SFR}}{R_{\mathrm{e}}^{\gamma/\beta}}\right)^{\beta},
\label{eq:eq10}
\end{equation}

where $\gamma/\beta\simeq1.25$ and $\beta = 0.91 \log(Z/Z_{\odot}) + 0.64$. Here, $L_{\mathrm{IR}}$ has been replaced by SFR, which provides a good approximation for massive, metal-rich SFGs (see Fig.~\ref{fig:figA1}). Notably, near solar metallicity, the exponent $\beta$ is broadly consistent with $1/n$, suggesting a similar underlying scaling despite the different parametrisations.

\begin{figure}
    \centering
    \includegraphics[width=0.8\columnwidth]{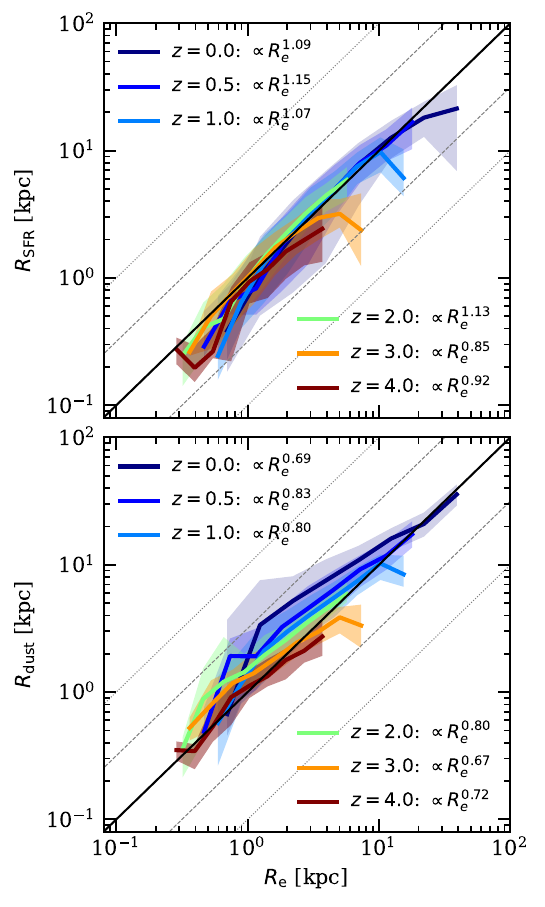}
    \caption{$R_\mathrm{SFR}$, the radius enclosing half of the stellar mass in stars younger than 10~Myr (top panel), and the half-mass dust radius (bottom panel) as a function of $R_\mathrm{e}$ at different redshifts. The legend indicates the best-fitting relation, $\propto R_\mathrm{e}^{f(z)}$. For reference, the observed relation of \citet{Qin2019} implies $R_\mathrm{IR}\propto R_\mathrm{e}^{0.6-0.75}$, and the TNG50 relation at solar metallicity $R_\mathrm{IR}\propto R_\mathrm{e}^{0.8}$ (Eq.~\ref{eq:eq11}).}
    \label{fig:fig5}
\end{figure}

Reconciling Eqs.~\ref{eq:eq9} and \ref{eq:eq10} requires a non-linear relation between the half-light radius and the spatial extent of the IR emission,

\begin{equation}
R_{\mathrm{IR}} \propto R_\mathrm{e}^{\gamma/(2\beta)},
\label{eq:eq11}
\end{equation}

\noindent i.e. $R_{\mathrm{IR}} \propto R_{\mathrm{e}}^{0.6-0.75}$ for the observed relation ($\gamma/\beta \simeq 1.25$--$1.5$), and $R_{\mathrm{IR}} \propto R_{\mathrm{e}}^{0.8}$ for the TNG50 relation at solar metallicity ($\gamma_{\mathrm{TNG}}/\beta_{\mathrm{TNG}} \approx 1.65$). $R_{\mathrm{IR}}$ can be approximated by the spatial distribution of dust, and therefore we adopt $R_{\mathrm{dust}}$, the half-mass radius of dust, as a proxy, and compare it with $R_{\mathrm{SFR}}$, the radius enclosing half of the stellar mass in stars younger than 10~Myr. Both quantities are shown in Fig.~\ref{fig:fig5} as a function of $R_{\mathrm{e}}$. At each redshift, we fit scaling relations of the form $\log R = \alpha(z) + f(z)\log R_{\mathrm{e}}$ for both $R_{\mathrm{SFR}}$ and $R_{\mathrm{dust}}$.

The size of the star-forming regions depends on redshift. At $z=3$ and 4, where the mass--size relation of the simulated galaxies becomes inverted (see Fig.~\ref{fig:figA2}), we find $f(z)=0.85$ and $0.92$, respectively: the star-forming regions are more compact than the stellar distribution. At $z\leq2$, $f(z)\approx1.1$, consistent with inside-out growth.

The dust, in contrast, is more extended than the stars at low redshift, and its size grows more slowly than $R_\mathrm{e}$ at all redshifts, with $f(z)=0.67$--$0.83$. This matches the exponent of $\approx0.8$ that Eq.~\ref{eq:eq11} predicts for the TNG50 relation at solar metallicity. In other words, the size dependence of the TNG50 relation is what one would expect if IRX traced the IR luminosity per unit area of the dust distribution.

The main difference from the observed relation is therefore not the dust geometry, but the exponent: $\beta_\mathrm{TNG}\approx0.34$ is about half of both the observed value near solar metallicity and the $1/n\approx0.7$ expected if IRX traced the dust surface density of star-forming gas. In other words, the IRX of TNG50-SFGs responds to changes in star formation surface density about half as strongly as observed, and this response does not increase with metallicity. Several aspects of the post-processing could contribute to this behaviour: dust is assigned to all cold gas within a region of $15\, R^\star_\mathrm{hm}$, so that part of the attenuating column is not associated with star-forming regions; the attenuation of the youngest stars ($<10$~Myr) is set by the birth-cloud model of \citet{Groves2008} rather than by the galaxy-scale dust distribution; and at low sSFR, older stars may contribute to dust heating, so that $L_\mathrm{IR}$ traces star formation less closely. Disentangling these contributions would require radiative transfer runs that vary the birth-cloud treatment and the spatial extent of the dust, which we defer to future work.

\subsection{Modelling and observational caveats}
\label{sec:caveats}

The offsets between TNG50-SFGs and the universal IRX relation reflect both the limitations of the simulated catalogues and their post-processing, and the range of galaxy properties over which the observed relation has been calibrated. We discuss both in the following paragraphs.

In this work, we assume a globally fixed ISM dust model for each galaxy, such that all gas cells share the same dust properties (MW-type dust; \citealt{Zubko2004}). In reality, galaxies are likely to host a mixture of dust types across different environments, which can significantly affect dust attenuation and, consequently, the inferred scaling relations \citep[see, e.g.,][]{Qin2024}. Furthermore, our assumption that dust content scales linearly with gas-phase metallicity may be oversimplified. For instance, \citet{Qin2024} show that, at higher metallicities, a larger fraction of dust resides in the diffuse ISM, leading to a decrease in the dust-to-stellar size ratio and resulting in deviations from models in which dust strictly traces star-forming gas. Such variations would change how much of the attenuating column is associated with star-forming regions, and could therefore affect not only the normalisation of IRX but also its response to the star formation surface density, i.e. the index $\beta$ discussed in Sect.~\ref{sec:interpretation}.

In addition, the DTM ratio is expected to vary both spatially within galaxies and as a function of cosmic time. For example, \citet{Vogelsberger2020} find a strong redshift evolution in this ratio by matching TNG dust-attenuated UV luminosity functions to observations, with values decreasing from $f_{\mathrm{dust}}\sim 0.9$ at $z=2$ to $f_{\mathrm{dust}}\sim 0.1$ at $z=6$. Such evolution significantly alters the predicted dust mass and consequently the redshift dependence of the IRX relation. In our post-processing, however, the DTM ratio acts mainly on the normalisation (see Sect.~\ref{sec:f_dust_var}). A DTM ratio higher than our fiducial value at $z\approx2$, as found by \citet{Vogelsberger2020}, would therefore reduce the negative offset at $z\geq2$, but higher values of $f_\mathrm{dust}$ flatten the $L_\mathrm{IR}$ dependence further rather than steepening it (Sect.~\ref{sec:z0}), and no choice of $f_\mathrm{dust}$ can move the TNG50-SFGs at $z\lesssim0.5$ into the luminosity range where the relation is calibrated. More recently, several studies have incorporated dust production, growth, and destruction directly into hydrodynamical simulations, with some also following the evolution of grain sizes, chemical composition, and extinction curves \citep{McKinnon2018,Hou2019,Li2019,Parente2022,Dubois2024,Trayford2026}. These approaches provide a more self-consistent treatment of dust evolution and offer a promising avenue for reducing the reliance on simplified assumptions about dust properties, such as a fixed DTM ratio, in post-processing.

Another source of uncertainty is the galaxy size. We derive $R_\mathrm{e}$ for TNG50-SFGs from the stellar half-mass radii and the light-to-mass size ratios of \citet{Suess2019a,Suess2019b}. This conversion may be biased by colour and stellar-age gradients and by dust attenuation, particularly in massive high-$z$ systems where these effects remain poorly constrained. Observed sizes are also measured differently: in the calibration sample of \citet{Qin2019}, $R_\mathrm{e}$ is the $r$-band semi-major-axis half-light radius from bulge+disc decompositions \citep{Simard2011}. Since $\gamma\approx1$ at the metallicities of TNG50-SFGs at $z\approx0$, a systematic difference of 0.1~dex in $R_\mathrm{e}$ translates into $\approx0.1$~dex in $\Delta\log\mathrm{IRX}$, so part of the size term found at matched $M_\ast$ and $L_\mathrm{IR}$ (Sect.~\ref{sec:z0}) may come from these different definitions.

Finally, the comparison depends on the range over which the universal IRX relation was calibrated. Its calibration sample consists of galaxies detected in both the IR (WISE) and the UV (GALEX), 97.5\% of which have $\log(L_\mathrm{IR}/L_\odot)\geq9.63$, and 97.5\% have $12+\log(\mathrm{O/H})\leq8.83$ (Sect.~\ref{sec:calibration}). Applying the relation to most TNG50-SFGs at $z\lesssim0.5$, which are fainter in the IR, therefore requires an extrapolation, and the offsets in this regime should be interpreted with this in mind. At higher redshift, the relation was validated with stacked samples, using metallicities and sizes inferred from scaling relations \citep{Qin2019}, so comparisons with individual simulated galaxies are less direct.

\section{Conclusions}
\label{sect:conclusions}

We analyse a sample of 7280 SFGs selected from TNG50 at six redshifts between $z=0$ and $z=4$, with dust attenuation computed by radiative transfer post-processing with \textsc{skirt}, adopting a MW-type dust model and a DTM ratio of $f_{\mathrm{dust}}=0.3$. We compare their IRX with the universal relation of \citet{Qin2019}, with its $z\approx0$ calibration sample, and with EAGLE-SFGs \citep{ManQiao2024}, and decompose the offset $\Delta\log\mathrm{IRX}$ into the contributions of the individual terms of the relation. Our main findings are summarised as follows:

\begin{itemize}

\item TNG50-SFGs are close to the universal IRX relation (within $\approx0.2$~dex) at $0.5\leq z\leq1$, but lie above it at $z=0$ ($\Delta\log\mathrm{IRX}\approx+0.5$~dex) and below it at $z\geq2$ ($\Delta\log\mathrm{IRX}\approx-0.25$~dex). Varying the DTM ratio between 0.1 and 1 changes IRX by 0.3--1.0~dex, mostly through $L_\mathrm{UV}$, but $L_\mathrm{IR}$ by at most 0.3~dex, so it shifts the overall normalisation without removing the redshift dependence of the offset. When individual galaxy properties are corrected (Sect.~\ref{sec:corrections}), gas-phase metallicity has the largest effect, as in observations \citep{Qin2019} and in EAGLE-SFGs \citep{ManQiao2024}.

\item TNG50-SFGs follow a well-defined relation of the same form as the observed one (mean residuals $\leq0.14$~dex and scatter of 0.2--0.35~dex; Sect.~\ref{sec:budget}), but with different indices. The IR luminosity index, $\beta_\mathrm{TNG}\approx0.34$, is independent of metallicity and about half of the observed value near solar metallicity, while $\gamma_\mathrm{TNG}$ increases with metallicity, so that $\gamma_\mathrm{TNG}/\beta_\mathrm{TNG}$ ranges from $\approx1.0$ to $\approx2.7$, compared with $\simeq1.25$ observed. The dust scales sub-linearly with the stellar size, $R_\mathrm{dust}\propto R_\mathrm{e}^{0.67-0.83}$, consistent with the size dependence of the TNG50 relation. The main difference from the observed relation is therefore the weak response of IRX to the star formation surface density, rather than the dust geometry.

\item The offset arises in two different regimes. At $z=0$, only 1\% (and $\approx 7\%$ at $z=0.5$) of TNG50-SFGs lie within the range of $L_\mathrm{IR}$ and metallicity where the relation is calibrated, and the offset is dominated by extrapolating the observed $L_\mathrm{IR}$ dependence to fainter galaxies, combined with the flatter $\beta_\mathrm{TNG}$ (the $L_\mathrm{IR}$ term contributes $+0.42$~dex at $z=0$). At $z\geq2$, most TNG50-SFGs (61--99\%) lie within the calibrated range, and the offset is a difference in normalisation ($-0.17$ to $-0.26$~dex), equivalent to a shift of $\approx0.2$~dex in the metallicity scale. The agreement at $z \approx 1$ results from a cancellation between these two contributions.

\item At $z\approx0$, TNG50-SFGs with $M_\ast>10^9 \, \mathrm{M_\odot}$ are $\approx1$~dex fainter in $L_\mathrm{IR}$ than the calibration sample at the same stellar mass, owing to both their lower SFRs and the IR selection of the observed sample. When galaxies are matched in both $M_\ast$ and $L_\mathrm{IR}$, their IRX agrees within 0.04~dex, and the remaining offset of $\approx+0.36$~dex in $\Delta\log\mathrm{IRX}$ comes from the size term, partly because $R_\mathrm{e}$ is defined differently in simulations and observations. EAGLE-SFGs show the same structure, and their agreement with the relation at $z\approx0$ results from a cancellation between terms.
\end{itemize}

Overall, the universal IRX relation is a powerful diagnostic of how simulations combine metal enrichment, star formation and galaxy structure, in particular within the range of $L_\mathrm{IR}$ and metallicity over which it has been calibrated. The main difference between TNG50 with standard post-processing and the observed relation is the weak dependence of IRX on the star formation surface density, while the normalisation of the SFMS and the gas-phase metallicities determine how this difference appears at each redshift. Further progress will benefit from observations of IR-faint galaxies, which would extend the range over which simulations can be compared with the relation, from consistent definitions of sizes and metallicities in simulations and observations, and from radiative transfer models that vary the birth-cloud treatment and the spatial extent of the dust.

\begin{acknowledgements}
PC acknowledges partial support from ANPCyT through grant PICT 2020-00582, and from the Tsung-Dao Lee Institute and Purple Mountain Observatory researcher visitor program. X.Z acknowledges support from the National Key Research and Development Program of China (2023YFA1608100), the National Science Foundation of China (12233005), and the China Manned Space Program with grant nos. CMS-CSST-2025-A08 and CMS-CSST-2025-A20, in part by the Office of Science and Technology, Shanghai Municipal Government (grant Nos. 24DX1400100, ZJ2023-ZD-001). A.K. has been supported by the National SKA Program of China (2025SKA0150104), the 100 talent program of the Sun Yat-sen University, and the Guangdong Basic and the Applied Basic Research Foundation with No. 2025A1515012670. M.Q. acknowledges support from the Natural Science Foundation of Shandong Province, China (Grant No. ZR2026QC2352Z). We thank the developers and maintainers of the following software used in this work: NumPy \citep{Harris2020}, SciPy \citep{Virtanen2020}, Matplotlib \citep{Hunter2007}, Astropy \citep{Astropy2013}, and Jupyter \citep{Kluyver2016}. 
\end{acknowledgements}

\section*{Data availability}

The data underlying this article will be shared on reasonable request to the corresponding authors. Most of the data of the IllustrisTNG project is, in fact, already openly available on the IllustrisTNG website, at \url{www.tng-project.org/data}. 

\bibliographystyle{aa}
\bibliography{biblio} 

\begin{appendix}
\onecolumn
\section{Scaling relations and corrections on IRX relation}
\label{sec:appendix}

\begin{figure*}[!htbp]
    \centering
    \includegraphics[width=\textwidth]{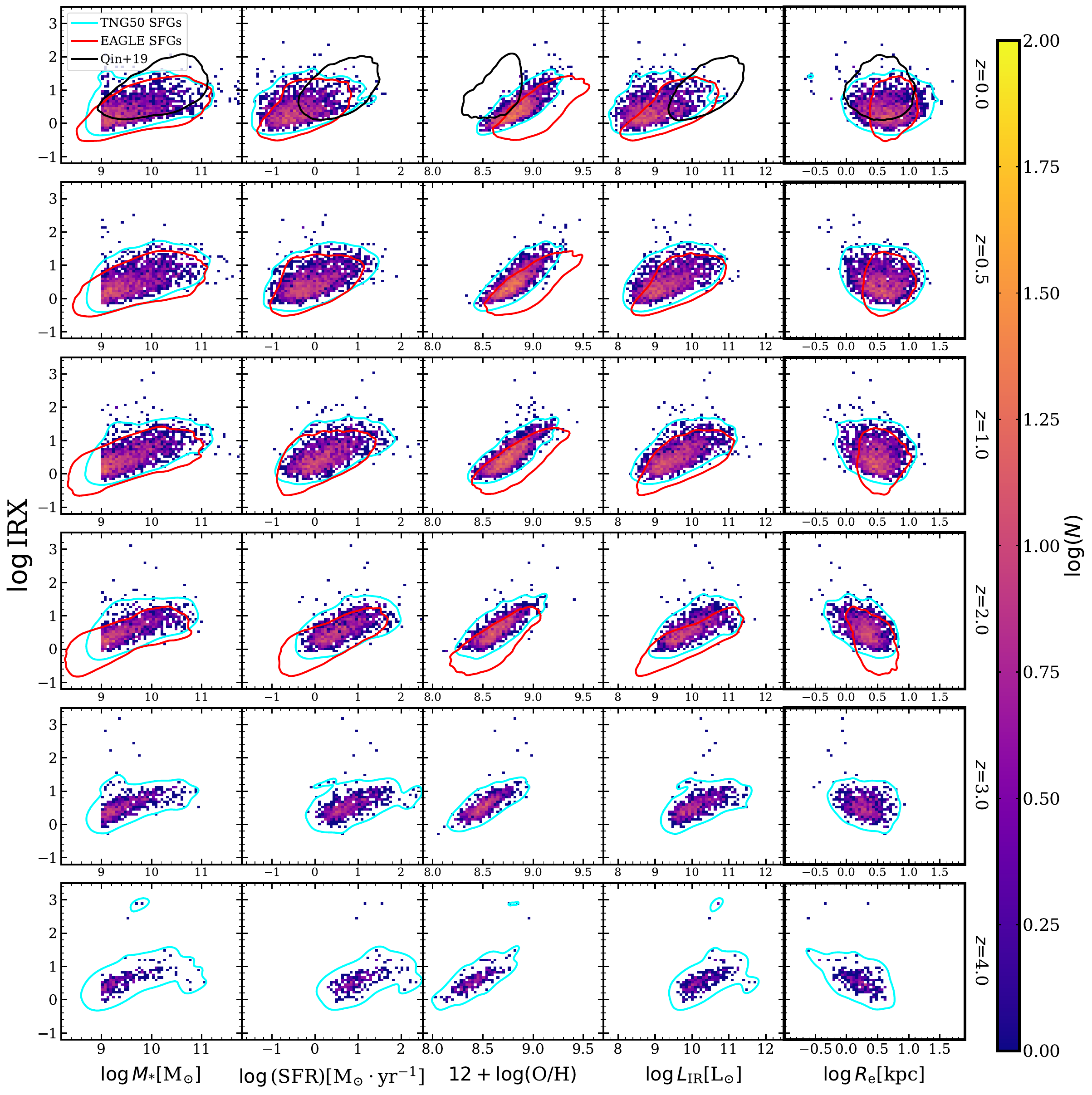}
    \caption{Relationships between $M_{\ast}$, SFR, metallicity, $L_{\mathrm{IR}}$, $R_{\mathrm{e}}$ and IRX for our sample TNG50-SFGs. Each row shows a different redshift from $z=0$ to $z=4$. The red and black contours show the EAGLE-SFG distribution from \citet{ManQiao2024} and the local SFGs from \citet{Qin2019}, respectively.}
    \label{fig:figA1}
\end{figure*}

\begin{figure*}
    \sidecaption
    \includegraphics[width=12cm]{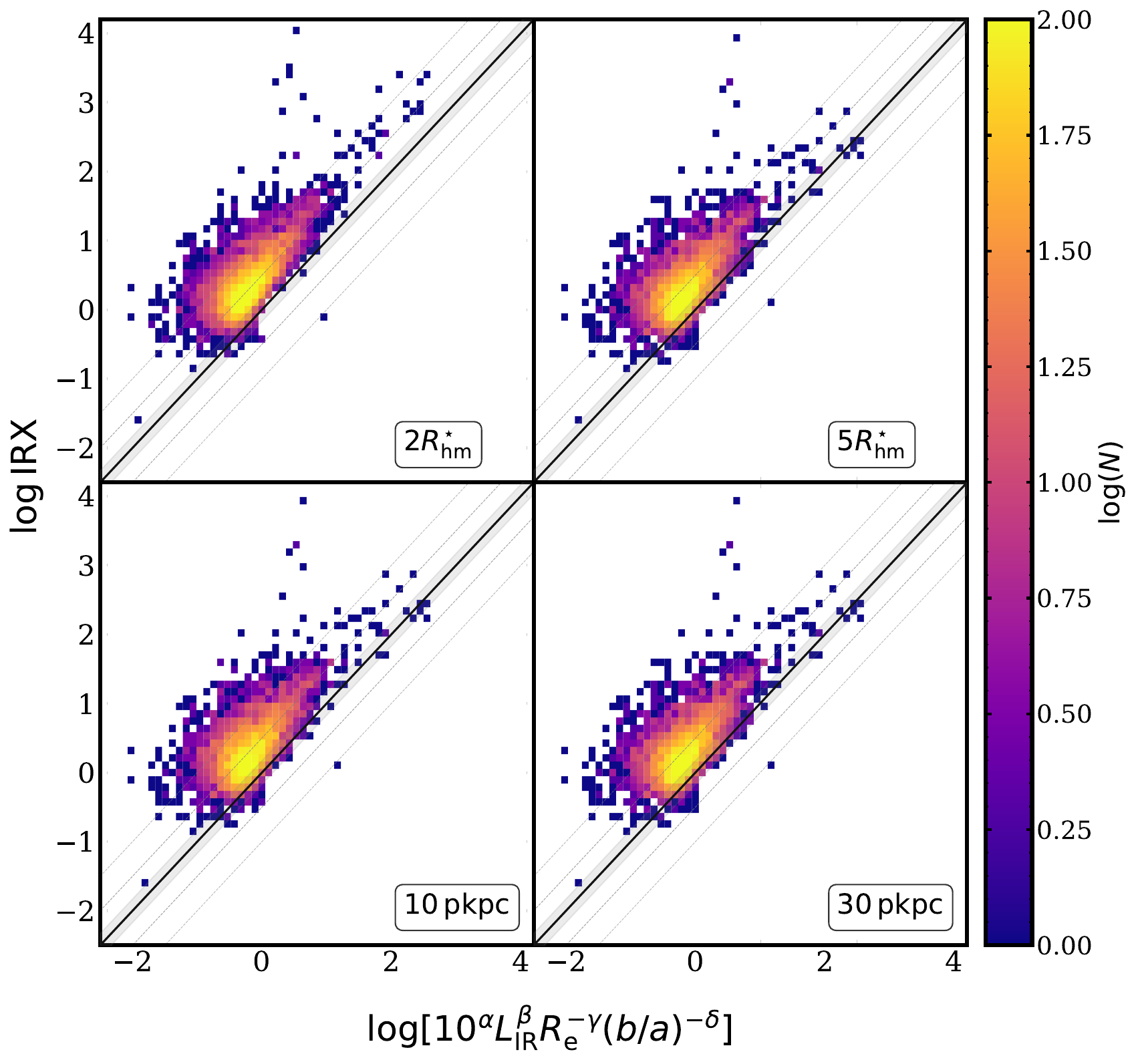}
    \caption{Same as Fig.~\ref{fig:fig2} but with \citet{Trcka2022} galaxy sample restricted to SFGs, shown here at $z=0$. Each subplot corresponds to a different spherical aperture within which the physical parameters entering Eq.~\ref{eq:eq3} are computed. The offset between the simulated and observed IRX remains largely unchanged across the different aperture definitions.}
    \label{fig:figTrcka} 
\end{figure*}

\begin{figure*}
    \sidecaption
    \includegraphics[width=12cm]{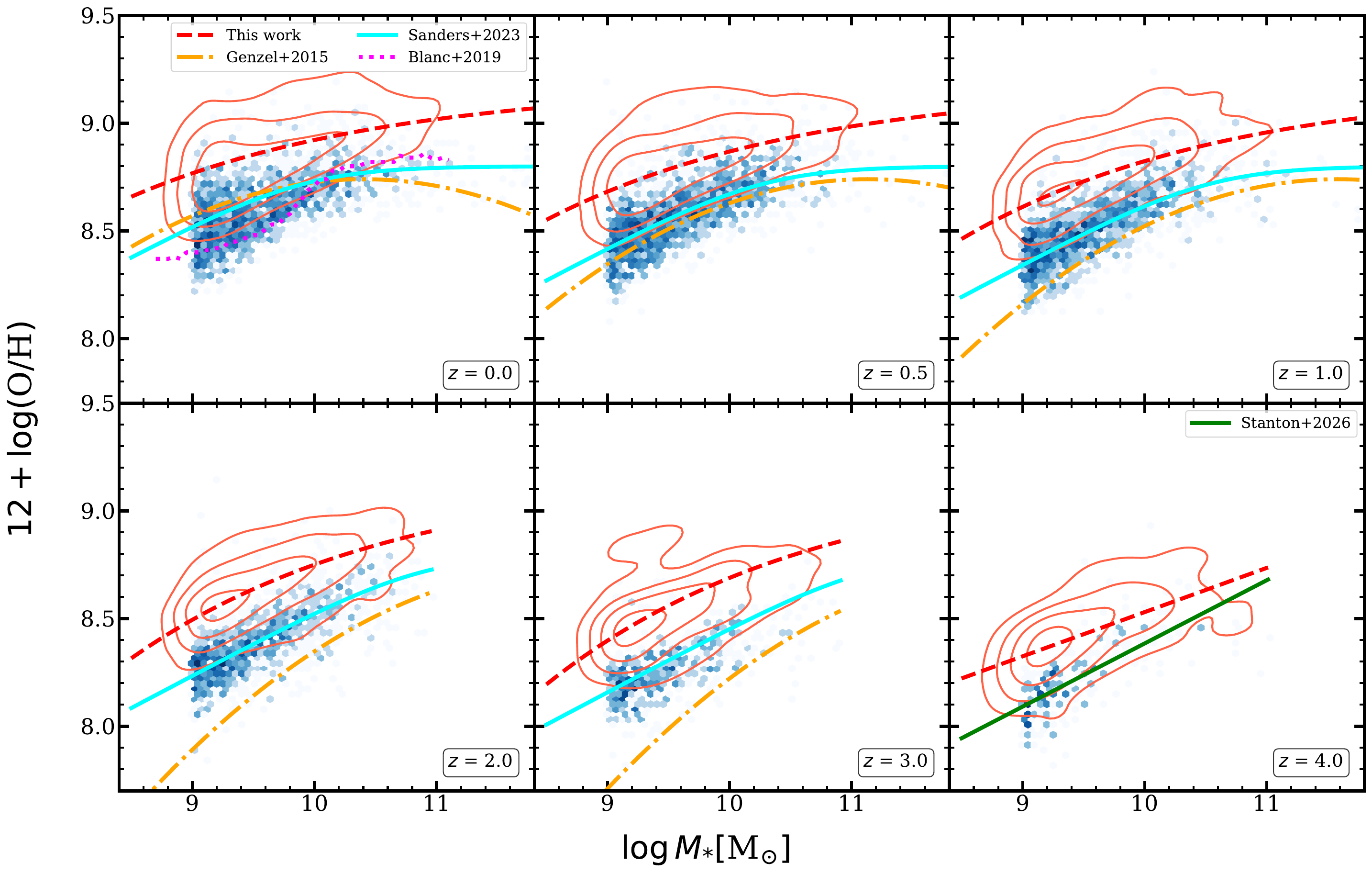}
    \caption{$M_{\ast}$--$Z$ relations of the TNG50-SFGs. The red contours show the distribution of the original stellar mass and metallicity, while the red lines represent the best-fitting relations using the same functional form adopted for the observed $M_{\ast}$--$Z$ scaling relation. The cyan lines show the $M_{\ast}$--$Z$ relations from \citet{Sanders2023}, and the green line at $z=4$ shows the relation from \citet{Stanton2026}. The hexagonal bins are colour-coded by the number of TNG50-SFGs corrected for the metallicity offset within each bin. The orange lines represent the $M_{\ast}$--$Z$ relations from \citet{Genzel2015}, while the magenta line shows the relation from \citet{Blanc2019}.}
    \label{fig:figA3}
\end{figure*}

\begin{figure*}
    \sidecaption
    \includegraphics[width=12cm]{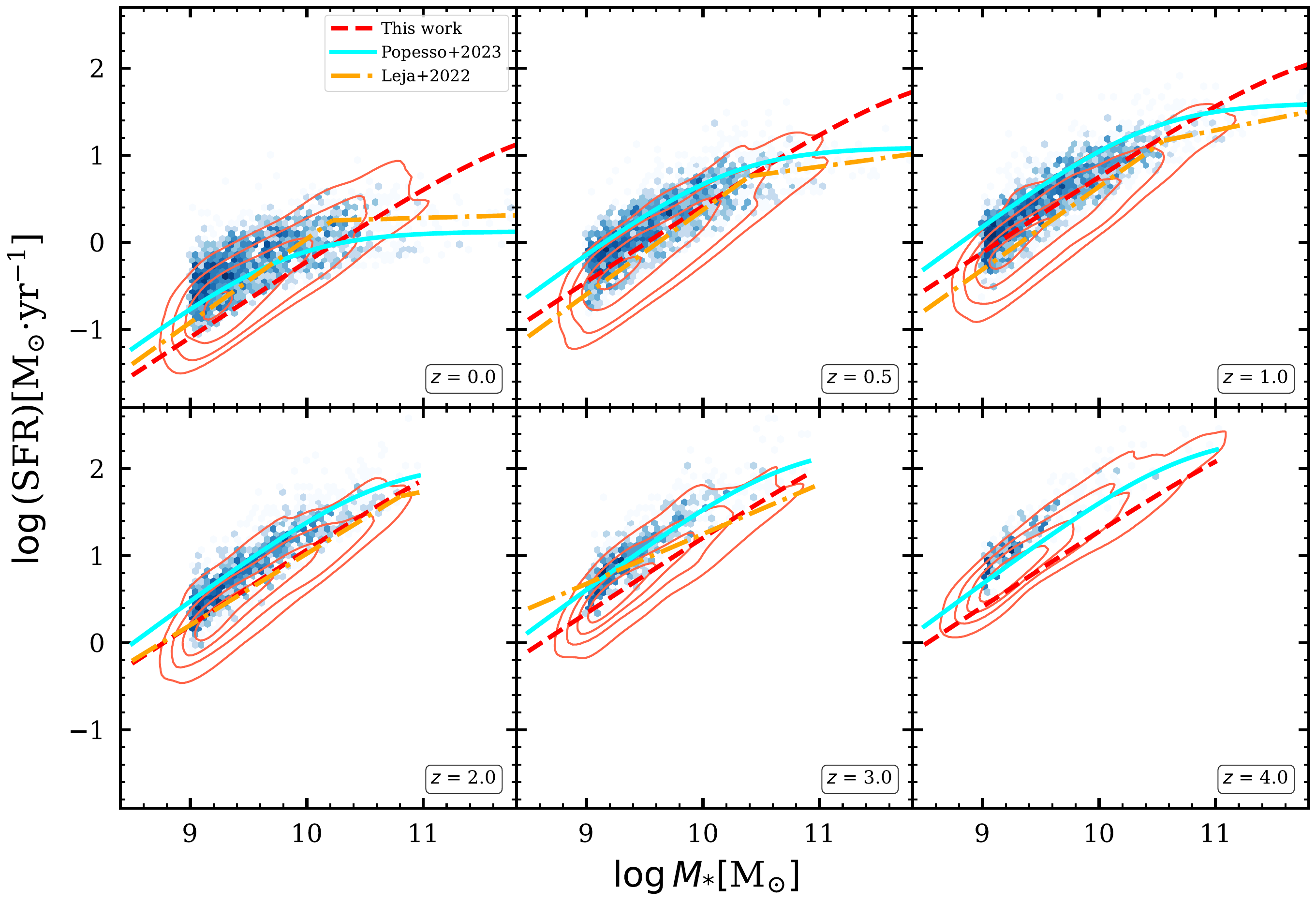}
    \caption{Comparison of the $M_{\ast}$--SFR for the TNG50-SFGs with original (red contours) and corrected (coloured hexagonal bins) SFR. The cyan lines show the $M_{\ast}$--SFR relations from \citet{Popesso2023}, whereas the red dashed lines show the best-fitting relations of the original TNG50-SFGs. The orange lines represent the $M_{\ast}$--SFR relations from \citet{Leja2022}.}
    \label{fig:figA4}
\end{figure*}

\begin{figure*}
    \sidecaption
    \includegraphics[width=12cm]{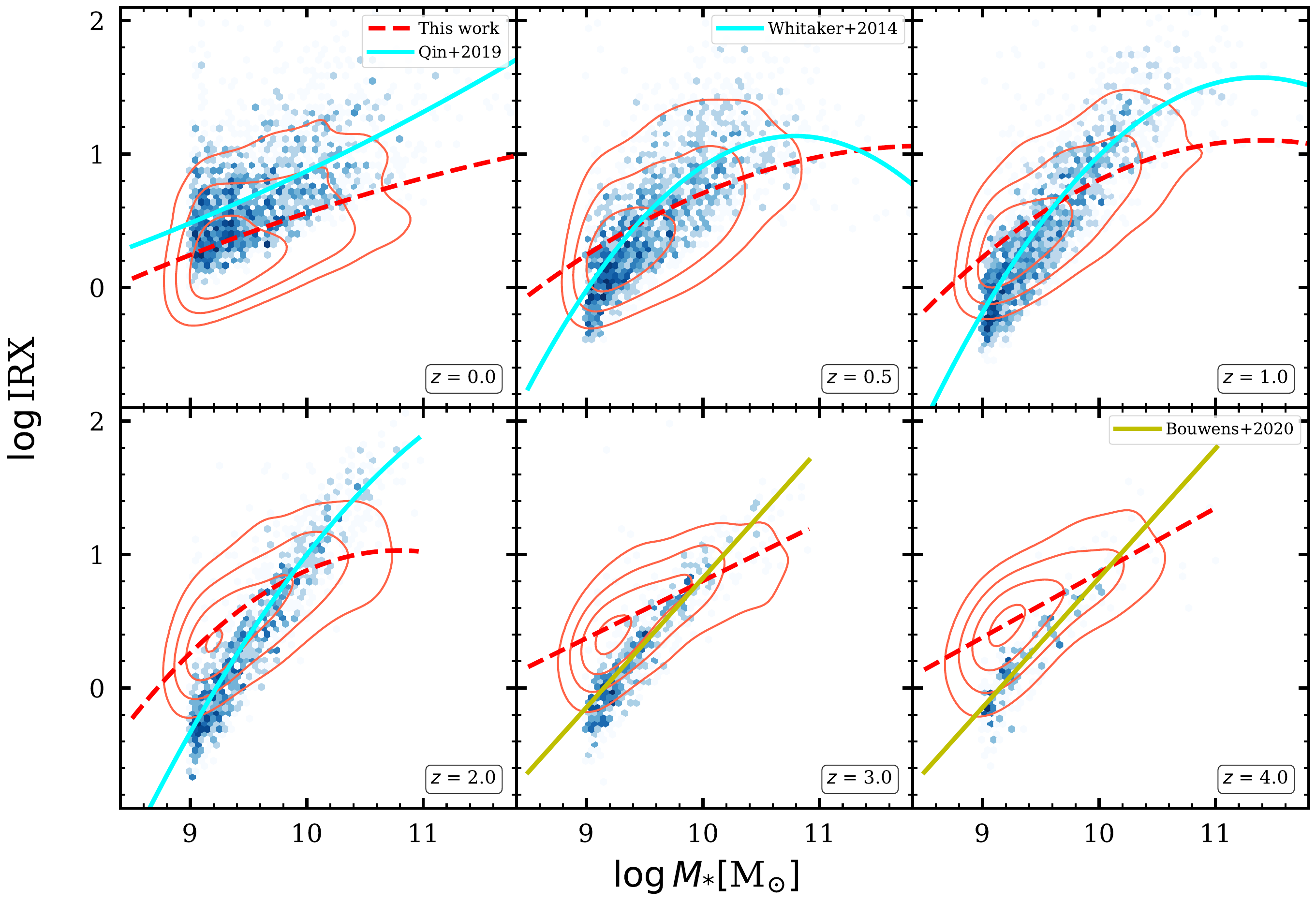}
    \caption{$M_{\ast}$--IRX relations for our original sample (red contours) and corrected sample (coloured hexagonal bins). The cyan lines show the observed $M_{\ast}$--IRX relations from \citet{Qin2019} at $z=0$ and \citet{Whitaker2014} at $0.5\leq z\leq 2$, while the green lines show the relation reported by \citet{Bouwens2020}. The red lines represent the best-fitting $M_{\ast}$--IRX relations of the TNG50-SFGs, obtained using the same functional form adopted for the corresponding observed scaling relations.}
    \label{fig:figA5}
\end{figure*}

\begin{figure*}
    \sidecaption
    \includegraphics[width=12cm]{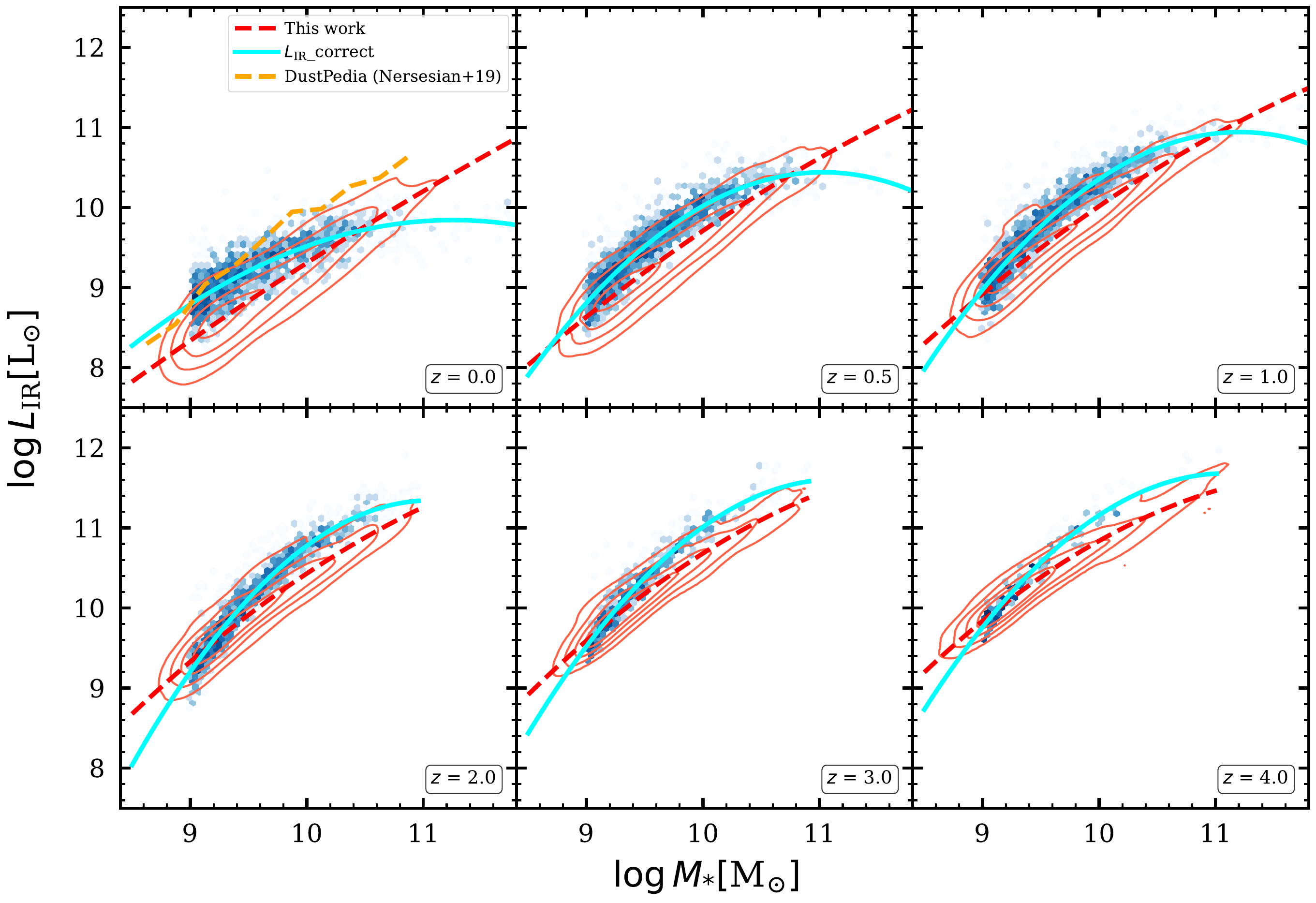}
    \caption{Comparison of the $M_{\ast}$--$L_{\mathrm{IR}}$ relations for the TNG50-SFGs with original (red contours) versus corrected (coloured hexagonal bins) $L_{\mathrm{IR}}$. The red and cyan lines mark the best-fitting $M_{\ast}$--$L_{\mathrm{IR}}$ relations as second-order polynomials in original and corrected $L_{\mathrm{IR}}$, respectively. In the $z=0$ panel, the orange dashed line shows the median $M_{\ast}$--$L_{\mathrm{IR}}$ relation of star-forming galaxies from \textsc{DustPedia} \citep{Nersesian2019}, selected with the same sSFR criterion as the simulated SFGs.}
    \label{fig:figA6}
\end{figure*}

\begin{figure*}
    \sidecaption
    \includegraphics[width=12cm]{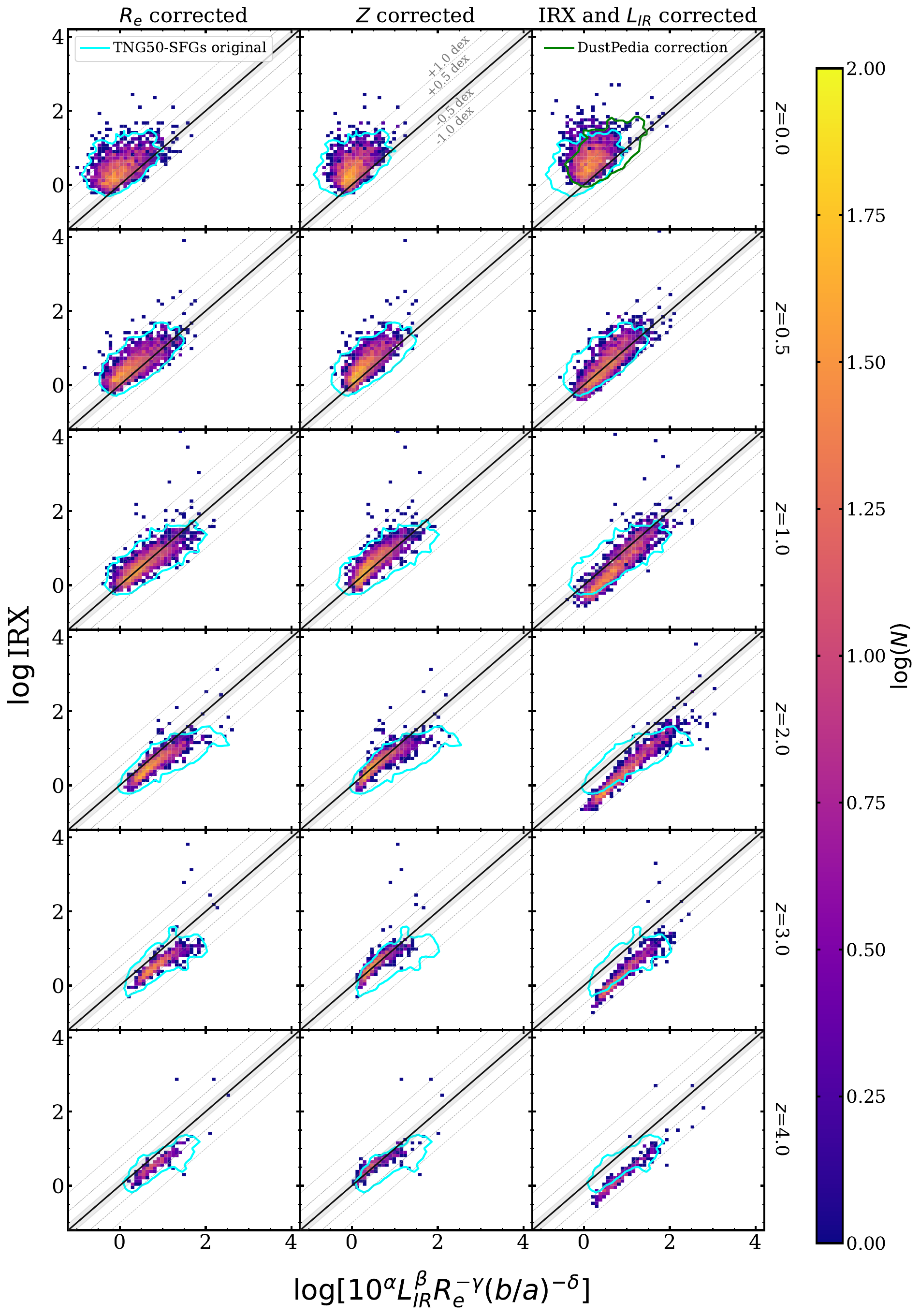}
    \caption{The predicted universal IRX relations for the TNG50-SFGs with corrected $R_{\mathrm{e}}$ (left column), $Z$ corrected (middle column), and with corrected $L_{\mathrm{IR}}$ and IRX (right column) at six redshifts representing each row. The cyan contour lines are the same as Fig.~\ref{fig:fig2}, with the original parameters. The IRX has better agreement with observations once metallicity is corrected, while the impact of the correction of $R_\mathrm{e}$ produces little change in the IRX, and the combination of the correction in $L_{\mathrm{IR}}$ and IRX produces a greater disagreement. For the $z=0$ panel, the green contours in the right column show the result obtained using the \textsc{DustPedia} correction. This alternative correction does not produce a substantial improvement in the agreement with the observed IRX relation.}
    \label{fig:figA7}
\end{figure*}

Figures \ref{fig:figA1} to \ref{fig:figA6} present additional results from our analysis that support our conclusions; further details are provided in the captions of each figure. Finally, in Fig.~\ref{fig:figA7}, we include the IRX-corrected relation for each physical parameter, adjusted according to their respective scaling relations.

In addition, Table~\ref{tab:sample_properties} summarises the median values alongside the minimum and maximum ranges for the main physical parameters ($\log M_{\ast}$, $\log \mathrm{SFR}$, $12+\log(\mathrm{O/H})$, $\log L_{\mathrm{IR}}$, and $\log R_{\mathrm{e}}$) of the TNG50, EAGLE, and \citet{Qin2019} SFGs across the studied redshift range.

\begin{table*}
\caption{Median and range [min, max] of the physical parameters across different redshift bins.}
\label{tab:sample_properties}
\centering
\small
\begin{tabular}{c c c c c c c c}
\hline\hline
$z$ & Sample & $N$ & $\log (M_{\ast}/\mathrm{M}_\odot)$ & $\log (\mathrm{SFR}/\mathrm{M}_\odot\,\mathrm{yr}^{-1})$ & $12 + \log(\mathrm{O/H})$ & $\log (L_{\mathrm{IR}}/\mathrm{L}_\odot)$ & $\log (R_{\mathrm{e}}/\mathrm{kpc})$ \\
\hline
0.00 & TNG50   & 1790  &  9.52 [ 9.00, 11.96] & $-0.45$ [$-1.38$, 1.61] & 8.84 [8.46, 9.76] &  8.88 [7.77, 11.18] &  0.64 [$-0.68$, 1.70] \\
0.08 & Qin+19  & 32354 & 10.18 [ 7.66, 11.36] &  0.55 [$-0.59$, 2.13] & 8.67 [8.09, 8.87] & 10.39 [9.12, 12.08] &  0.62 [$-2.00$, 1.41] \\
0.10 & EAGLE   & 7867  &  9.68 [ 8.50, 11.51] & $-0.35$ [$-1.52$, 1.31] & 9.05 [8.40, 9.66] &  9.38 [7.61, 11.07] &  0.70 [ 0.16, 1.41] \\
\hline
0.50 & TNG50   & 1767  &  9.57 [ 9.00, 11.83] & $-0.09$ [$-1.11$, 1.86] & 8.81 [8.33, 9.46] &  9.27 [8.23, 11.27] &  0.56 [$-0.44$, 1.34] \\
     & EAGLE   & 11609 &  9.52 [ 8.50, 11.60] & $-0.18$ [$-1.31$, 1.55] & 8.95 [8.19, 9.59] &  9.50 [8.17, 11.21] &  0.59 [ 0.11, 1.39] \\
\hline
1.00 & TNG50   & 1710  &  9.53 [ 9.00, 11.97] &  0.19 [$-0.92$, 2.40] & 8.76 [8.35, 9.43] &  9.55 [8.35, 11.81] &  0.47 [$-0.33$, 1.27] \\
     & EAGLE   & 15867 &  9.28 [ 8.50, 11.52] & $-0.08$ [$-1.06$, 1.74] & 8.82 [8.18, 9.66] &  9.57 [8.31, 11.48] &  0.49 [$-0.04$, 1.26] \\
\hline
2.00 & TNG50   & 1149  &  9.48 [ 9.00, 10.97] &  0.57 [$-0.35$, 2.44] & 8.63 [8.10, 9.39] &  9.92 [8.93, 11.71] &  0.31 [$-0.58$, 0.83] \\
     & EAGLE   & 15070 &  9.01 [ 8.50, 11.27] &  0.02 [$-0.89$, 2.00] & 8.58 [7.98, 9.45] &  9.64 [8.17, 11.74] &  0.42 [$-0.22$, 1.27] \\
\hline
3.00 & TNG50   & 620   &  9.40 [ 9.00, 10.91] &  0.74 [$-0.14$, 2.40] & 8.51 [8.07, 8.98] & 10.09 [9.41, 11.50] &  0.29 [$-0.53$, 0.94] \\
\hline
4.00 & TNG50   & 244   &  9.33 [ 9.00, 11.02] &  0.95 [ 0.29, 2.36] & 8.41 [8.03, 9.07] & 10.25 [9.64, 11.72] &  0.21 [$-0.62$, 0.63] \\
\hline
\end{tabular}
\tablefoot{The listed physical parameters ($\log M_{\ast}$, $\log \mathrm{SFR}$, $12+\log(\mathrm{O/H})$, $\log L_{\mathrm{IR}}$, and $\log R_{\mathrm{e}}$) correspond to the redshift bins shown in Fig.~\ref{fig:figA1}. TNG50 and EAGLE SFGs are evaluated at their respective snapshot redshifts, while the \citet{Qin2019} local sample is reported at its median redshift ($0.04 < z < 0.15$).}
\end{table*}

\end{appendix}
\end{document}